**On the microscopic origin of reversible and irreversible reactions of LiNi$_x$Co$_y$Mn$_x$ cathode materials: Ni-O hybrid bond formation vs. cationic and anionic redox**

*Karin Kleiner*, Claire A. Murray, Cristina Grosu, Sarah J. Day, Martin Winter, Peter Nagel, Stefan Schuppler, Michael Merz*

Dr. Karin Kleiner, Prof. Martin Winter
Münster Electrochemical Energy Technology, University of Muenster (MEET, WWU)
Corrensstr. 46,
48149 Muenster, Germany
E-Mail: karin.kleiner@wwu.de

Dr. Claire Murray, Dr. Sarah Day
Beamline I11, Diamond Light Source (I11, DLS)
Harwell Science and Innovation Campus, Diamond House
Didcot OX11 0DE

Cristina Grosu
Institute of Energy and Climate Research, Research Center Jülich (IEK9, FZJ)
52425 Juelich, Germany

Prof. Martin Winter
Helmholtz-Institute Münster, Forschungszentrum Jülich GmbH
Corrensstr. 46,
48149 Muenster, Germany

Dr. Peter Nagel, Dr. Stefan Schuppler, Dr. Michael Merz
Institute for Quantum Materials and Technologies, Karlsruhe Institute of Technology (IQMT, KIT)
Hermann-von-Helmholtz-Platz 1
76344 Eggenstein-Leopoldshafen, Germany

Energy density limitations of layered oxides with different Ni contents, i.e., of the conventional cathode materials in Li-ion batteries, are investigated across the first discharge cycle using advanced spectroscopy and state-of-the-art diffraction. For the first time unambiguous experimental evidence is provided, that redox reactions in NCMs proceed *via* a reversible oxidation of Ni and a hybridization with O, and not, as widely assumed, *via* pure cationic or more recently discussed, pure anionic redox processes. Once Ni-O hybrid states are formed, the sites cannot be further oxidized. Instead, irreversible reactions set in which lead to a structural collapse and thus, the lack of ionic Ni limits the reversible capacity. Moreover, the degree of hybridization, which varies with the Ni content, triggers the electronic structure and the operation potential of the cathodes. With an increasing amount of Ni, the covalent character of the materials increases and the potential decreases.

## 1. Introduction

The heavy reliance of the transport industry on fossil fuels means it is currently producing more than 30% of greenhouse gas emissions.[1,2] Reaching global emission targets therefore requires a widespread adoption of environmental sustainably, electric drive technologies. Electric vehicles cannot only mitigate pollution issues from traffic, greenhouse gas emissions would also become limited to the manufacturing process which allows for large-scale treatment. Li-ion batteries, ubiquitous in consumer electronics nowadays, play a key role in achieving a widespread adoption of electric vehicles due to their relatively high energy densities and compact design.[3] However, for a successful market penetration even higher energy densities (capacity × discharge voltage) and longer lifetimes are required.[4–6] The limiting component in current generations of Li-ion batteries is the cathode material.[4,7]

The most commercially used and promising future cathode materials belong to the class of rhombohedral, layered oxides (Li*Me*O$_2$, *Me* = Ni, Co, Mn, Li, Al, Mg) or structurally related materials, in which transition metals, Mg or Al and Li-ions occupy edge-sharing oxygen octahedra in alternating layers.[4–8] A large number of these materials exist and Ni- and Li-rich layered oxides show the highest energy densities.[4,9] Cathode materials undergo reversible electron exchange (redox) reactions when they have to release and accommodate Li ions upon charge and discharge. This is challenging from a materials perspective due to the associated large crystallographic and electronic changes that lead to a limitation in energy density and to severe degradation reactions, in particular for layered oxides.[10–12] While changes in the crystallographic structure are relatively easy to study and are consequently well understood, [10,13–16] the electronic properties of these materials and their changes upon operation are rarely investigated due to the need of either soft x-ray absorption or emission spectroscopy, for which the techniques require ultra-high vacuum, and data analysis is not as far developed.[17–23] This leads to an incomplete understanding of the electrochemical processes determining the observed voltage, the reversible capacity and the limitations in the cycling stability of layered oxides. Tarascon et al. showed early on that the charge compensation is much more complex than the widely accepted cationic redox process ($Me^{2+ \text{ or } 3+} \rightleftharpoons Me^{4+} + 1$ or $2$ e$^-$) indicates [24–26]: Based on changes in the $Me$-O distance upon charge and discharge he suggested that the reversible process might include hybridizations of $Me$ $3d$ and O $2p$ states (covalence).[27]

The present work re-raises the intriguing question about the redox process in layered oxides and the covalence of $Me$-O bonds while putting emphasis on its importance in understanding capacity and voltage limitations. It also complements more recent studies claiming that oxygen is redox active (O$^{2-} \rightleftharpoons$ O$^{(-2+x)-}$ + x e$^-$, 0<x<1, so-called 'anionic redox'), especially at high states of charge.[26,28–34] The capacity and the open cell voltage are determined by the redox active centres and their valence states [35,36], which are investigated with near edge X-ray absorption fine structure (NEXAFS) at the (transition) metal (Me) $L_{2,3}$ and oxygen (O) K edges. The results give not only unique insights into oxidation and reduction processes in Ni containing layered oxides, but also allow to study changes in the electronic nature of $Me$ and O valence states (*e.g.* the ligand character of $Me$ $3d$ states and thus the covalence of $Me$-O bonds). Redox processes at transition metal sites are always affiliated with a change in $Me$-O hybridization, ruling out that pure cationic or anionic redox processes occur at any point upon cycling. The combination of the NEXAFS results with *operando* synchrotron powder diffraction (SXPD) enables the exploration of stability limits and thus provides important information on how to increase the energy density and lifetime of present Li-ion batteries.

## 2. Results and Discussion

Rhombohedral Li transition metal oxides (Li$Me$O$_2$, Me = Ni, Co, Mn) such as NCM 111 (LiNi$_{1/3}$Co$_{1/3}$Mn$_{1/3}$O$_2$), NCM 622 (LiNi$_{0.6}$Co$_{0.2}$Mn$_{0.2}$O$_2$) and NCM 811 (LiNi$_{0.8}$Co$_{0.1}$Mn$_{0.1}$O$_2$) are suitable for highly reversible Li de-/intercalation and redox reaction according to the assumed chemical reaction given in **Equation 1**.

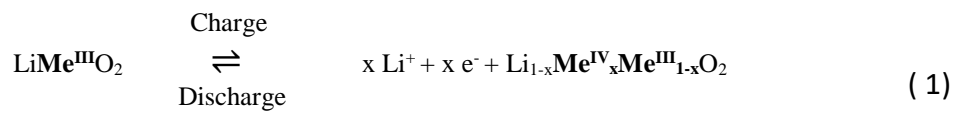

$$\text{Li}\textbf{Me}^{\textbf{III}}\text{O}_2 \underset{\text{Discharge}}{\overset{\text{Charge}}{\rightleftharpoons}} \text{x Li}^+ + \text{x e}^- + \text{Li}_{1\text{-x}}\textbf{Me}^{\textbf{IV}}_{\textbf{x}}\textbf{Me}^{\textbf{III}}_{\textbf{1-x}}\text{O}_2 \qquad (1)$$

The electrochemical performance of NCMs in this study is shown in **Figure 1** (for details about the cycling procedure see experimental section). While NCM 811 and NCM 622 show a significantly higher capacity than NCM 111 (**Figure 1A**), the rate capability of the materials is very similar. However, NCM 811 suffers from a faster capacity decay starting after 25 cycles if compared to NCM 111 and NCM 622. The dq/dV plots of the first cycle, for which the peaks can be assigned to redox processes, are given in **Figure 1B**. Reference dq/dV plots of LiNiO$_2$ (LNO), LiCoO$_2$ (LCO) and Li$_2$MnO$_3$ (LLO) (dashed lines in **Figure 1B**) reveal, that redox processes observed in NCMs occur at similar voltages as it is the case for LNO. The redox processes in LCO appear at significantly higher voltages while LLO does not show any significant redox activity (the choice of reference material is discussed in **S1**). NEXAFS spectra were taken upon the first discharge of the materials (**Figure 1C,** blue circles 1-5). The spectra are shown in **Figure 2** (Ni $L_{2,3}$ and O K edges, also labelled with the blue circles 1-5) and **S2** (Mn $L_{2,3}$ and Co $L_{2,3}$ edges).

NEXAFS is a powerful technique to study the electronic structure of materials if analyzed on edges allowing core electrons to be excited into valence orbitals following dipole selection rules.[17] In first order allowed transitions do correspond to the type $Me$ 2p$^6$ 3d$^n$ → $Me$ 2p$^5$ 3d$^{n+1}$ (*Me L<sub>2,3</sub>*) at the transition metal sites, while *Me*-O hybridizations are included by admixing transitions of the type $Me\ 2p^6\ 3d^{n+1}\ O\ 2p^5 \rightarrow Me\ 2p^5\ 3d^{n+2}\ O\ 2p^5$, where a ligand-to-metal charge-transfer leads to holes at oxygen sites,  and $O\ 1s^2\ O\ 2p^x \rightarrow O\ 1s^1\ O\ 2p^{x+1}$ (O K) at the ligand sites. Therefore NEXAFS at the *Me L<sub>2,3</sub>* and O K edge provides useful information about the oxidation states of 3d metals and O as well as about the electronic nature of the *Me*-O bonds.[17] Ni and O are the main redox active elements in the present samples, since the Ni *L<sub>2,3</sub>* and O K NEXAFS spectra, recorded in fluorescence yield (FY) with a sampling depth of ~100 nm, show obvious differences between the charged and the discharged states (**Figure 2**). However, in contrast to literature [24,26,37,38], the Co and Mn *L<sub>2,3</sub>* spectra do not reveal any significant differences between the two states of charge (see **S2**). Reasons for the deviation from literature include: (i) Complementary studies are often performed at higher excitation energies (transition metal K edges) due to a much easier experimental setup  that does not require ultra-high vacuum and an inert sample transfer as in this study.[24] In this case, excitations into the valence states of layered oxides are quadrupole transitions only allowed due to distortions of the *Me*-O octahedra or/and vibrations of the atoms. Consequently, the peaks are relatively weak and *Me* 3d-O 2p interactions are often neglected upon data analysis; (ii) The surface of the NCM particles degrades much faster than the bulk, and thus surface sensitive measurements differ significantly from bulk measurements [38]; (iii) A significantly higher end of charge voltage leads to states of charge above the stability window of the materials, thus forcing irreversible redox reactions.[26,34]

Upon discharge, low energy shoulders of the Ni *L<sub>3</sub>* (853 eV) and *L<sub>2</sub>* edge (871 eV) appear (**Figure 2A**, dashed lines), while O K edge peaks at 529 eV and 530 eV, which are assigned to excitations into O 2p states [39,40], are decreasing (**Figure 2B**, dashed lines). The *Me L<sub>2,3</sub>* edges were simulated with charge-transfer multiplet (CTM) calculations in order to extract changes in the valence states of the transition metals in the NCMs upon discharge.[22,23] A detailed description of the theoretical background is given in **S3**. **Figure 3** is an overview of the CTM calculations of $Ni^{2+}$, $Ni^{3+}$ and $Ni^{4+}$. More details about the simulations of the *Me* $L_{2,3}$ edges (*Me* = Ni, Co, Mn) are provided in **S4** For the calculation of the Ni $L_{2,3}$ spectra, the ligand-field (LF) energy was varied from 1 eV to 4 eV and the ligand-to-metal charge-transfer (LMCT) energy $\Delta$, which determines the covalence of the *Me*-O host structure, from -15 eV to 15 eV. The core hole potential $U_{pd}$ and the on-site Coulomb interactions $U_{dd}$ were kept constant ($U_{pd}$ - $U_{dd}$ = 1 eV). Significant spectral changes upon variation of $\Delta$ and LMCT are evident from **Figure 3**. A good match of the charged NCM spectra and the simulations is obtained for a very covalent $Ni^{3+}$ configuration (23% $3d^7$ $2p^6$ ⇌ 77% $3d^8$ $2p^5$, LF energy = 3 eV, **Figure 4A-C**). Upon discharge the peaks, which belong to this configuration are still present (shifted to slightly lower energies) but new peaks at 853 eV and 871 eV appear (**Figure 2A**, dashed lines). These peaks are assigned to a very ionic $Ni^{2+}$ configuration (< 96% $3d^8$ $2p^6$ ⇌ > 4% $3d^9$ $2p^5$). The $Ni^{2+}/$ $Ni^{3+}$ ratio increases upon discharge from < 4% to 42% for NCM 111, 38% for NCM 622 and 31% for NCM 811 (**Figure 4D-F**). This means that the electronic structure of the NCMs changes from a very covalent (> 96% of covalent $Ni^{3+}$) in the charged to a more ionic character (58%-69% of covalent $Ni^{3+}$, 42%-31% of ionic $Ni^{2+}$) in the discharged state. These changes alter the band gap (the valence states) and the electronic structure of the NCMs change from a charge-transfer type (covalent) to a more Mott-Hubbard type (more ionic) semiconductor, see **S3** According to **Equation (** 1, $Ni^{4+}$ is the expected oxidation state of Ni in charged NCMs. However, none of the calculated $Ni^{4+}$ configurations show a good match with the measured data. Moreover, it is well known that $Ni^{3+}$ has a negative charge-transfer character which stabilizes the ligand hole character (the $3d^8$ $2p^5$ configuration) and therefore makes the formation of $Ni^{4+}$ practically impossible.[41] The presence of $Ni^{2+}$ in the discharged state can partially be explained by the 4+ configuration of the more electronegative Mn as observed spectroscopically at the Mn $L_{2,3}$ edge, see **S4**. However, in NCM 811 and NCM 622, more $Ni^{2+}$ is present than expected from the $Mn^{4+}$ content which might arise from a small oxygen deficiency ($< 0.2\%$) as discussed already in [23].

The hybridization of *Me* 3d and O 2p orbitals leads to holes at O 2p sites (*e.g.* Ni $3d^{7+x}$ O $2p^{6-x}$, $x > 0$) which can be detected using O K edge XAS.[37,39,40] Such a hole at ligand sites thus directly probes the ligand character of the valence states and therefore the covalence of the *Me*-O interactions.[42] The amount of holes in the O 2p orbitals decreases upon discharge (peaks at 529 eV and 530 eV, **Figure 2B**, dashed lines) due to the decrease in the covalent character of the *Me*-O bonds, consistent with the Ni $L_{2,3}$ NEXAFS analysis. Comparing the O K edge NEXAFS spectra (**Figure 5**) of charged (dashed lines) and discharged (solid lines) NCM 111, NCM 622, and NCM 811 to O K reference spectra of LNO ($LiNiO_2$), LLO ($Li[Li_{1/3}Mn_{2/3}]O_2$) and LCO ($LiCoO_2$) helps to distinguish the individual contributions of Ni-, Co- and Mn-O hybridizations. The peaks A-E (**Figure 5**) are attributed to O 1s (O K shell) core electron excitations into empty O 2p states.[39,40] The hybridization of Ni and O orbitals leads to peaks at 529 eV (peak A) and 530 eV (peak C) as evidenced by the pre peaks in the O K spectra of discharged LNO. One Co-O peak in the discharged LCO spectrum is found at 531 eV (peak D) while electronic holes from the Mn-O hybridization (discharged LLO) lead to peaks at 530 eV and 532.5 eV (peak B and E). The peaks at energies around 534 eV (peak F and G) are assigned to electron excitations into Hubbard bands [37] and the broader peaks above 535 eV to transitions into hybridized states of O 2p and Me 4s or higher unoccupied Me orbitals.[37,40,43] The changes in the O K NCM spectra upon charge are comparable to changes observed for LNO (**Figure 5**, line I and II and **S5**), confirming that mainly Ni contributes to the charge compensation of layered oxides which is in agreement with the results from the dq/dV plots and the Ni $L_{2,3}$ NEXAFS analysis.

The above discussed results question long-standing assumptions about the charge compensation mechanism in NCMs. The formal and widely accepted redox process is the cationic oxidation/reduction $Me^{2+\,or\,3+} \rightleftharpoons Me^{4+} + 1$ or 2 e⁻ (with *Me* = Ni, Co), schematically depicted in **Figure 6A** (middle panel).[24–26] However, the present results rather indicate that the reversible charge compensation only takes place at Ni-O sites and Ni does not reach the expected +4 oxidation state due to its strong negative charge-transfer character. The probability of finding electron holes at O is much higher than at $Ni^{3+}$ meaning that $Ni^{3+}$ is stabilized by electron density shifted from oxygen ligands to Ni upon charge. As a first approximation, the number of electrons around Ni does not differ significantly from $Ni^{2+}$, which is the chemically stable configuration. These findings also contradict claims that O is redox active in the *Me*-O host structures at high states of charge ('anionic redox') [33,44], depicted in **Figure 6A** (right side), since the electron density at oxygen sites decreases due to the formation of covalent bonds with its transition metal neighbours (**Figure 6B**, middle panel) and not due to the redox process $O^{2-} \rightleftharpoons O^{(-2+x)-} + x\ e^{-} (0<x<1)$. When Tarascon et al. shaped the term anionic redox, it described redox processes affiliated with *Me*-O hybrid orbitals, a depiction which is very similar to the discussed Ni-O redox activity in NCMs.[27–29,31,32] These findings are partially supported by recent advances in absorption spectroscopy, although pure cationic and anionic redox processes are still discussed for voltages > 4.3 V (applied end of charge voltage in the present work), where Co and Mn might become redox active.[26,34] However, O 2p holes in $LiNiO_2$, $LiCoO_2$ and $Li_2MnO_3$ (**Figure 5**, peaks A-E) show, that all transition metals in $LiMeO_2$ form covalent *Me*-O bonds and thus the charge compensation will always affect *Me* and O at the same time, even if this means that such reactions are affiliated with oxygen release, *Me* dissolution and/or phase transformations.[10,12,27,45–48] The present work goes even further showing that the reversible charge compensation in NCMs changes the electronic structure from an ionic towards a more covalent semiconductor *via* changes in the hybridization of the Ni-O bonds. In combination with the semi-quantitative NEXAFS analysis these results enable for the first time a discussion about the microscopic origin of capacity and electrode potential limitations in NCMs, up to now empirical determined values, although hints to their physical origin were reported by Goodenough et al. early on.[35,49]

According to **Equation (** 1, theoretically one mole of Li-ions and one mole of electrons can be extracted per mole Li$Me$O$_2$ ($Me$ =Ni, Co, Mn). However, in reality the cathode materials undergo irreversible decomposition reactions at low Li contents ($x$ in Li$_x$$Me$O$_2$ < 0.5). Desired and detrimental reactions are highly correlated to crystallographic changes and electronic properties of the materials.[23] The interstitial sites upon Li-diffusion are the tetrahedral sites in the Li layer and the edge length of these tetrahedra (O-O (tet.) in **Figure 7A**), in turn, defines the activation barrier for Li diffusion.[50,51] The smaller the tetrahedral, the larger the attractive interactions between Li and O and thus the higher the activation barriers for the diffusion process. With an increasing state of charge (from the right to the left, **Figure 7A**) repulsive interactions between negatively charged, opposing oxygen layers increase as Li-ions, the positive charges placed between the layers, are extracted. This leads to an increasing size of the tetrahedral sites and thus to a decrease in the activation barrier. At the same time, the $Me$-O distances decrease which is consistent with the increasingly covalent character of the $Me$-O host structure, **Figure 7B**. At ca. 150 mAh/g the size of the tetrahedra, and thus the O-O (tet.) distance (**Figure 7A**), reaches a maximum. At higher states of charge, the decreasing edge length of the tetrahedral sites leads to a breakdown in diffusion kinetics and irreversible reactions like oxygen release, phase transformation and transition metal dissolution are observed (**Figure 6B**, right panel).[10,12,45–48] At this point, the amount of Ni$^{2+}$ (**Figure 7C**) becomes almost zero (< 0.02 per Li$Me$O$_2$) which means that the Ni$^{2+}$ content present in the discharged NCMs limits the reversible capacity. This strongly suggests that if there is a way to increase the available amount of Ni$^{2+}$ in discharged NCMs, *e.g.* by substituting O with a stronger electron donating anion, the reversible capacity (which is determined by the point where Ni$^{2+}$ is suppressed to zero) would increase along with it. Note that the collapse of the tetrahedral sites is much stronger in the case of NCM 811 compared to NCM 622 and NCM 111 at the end of charge (**Figure 7B**), which can be related to the covalent character of the host structure. From NCM 111 *via* NCM 622 to NCM 811 the covalent Ni$^{3+}$ content increases from 14% to 37% and finally to 55% in the discharged NCMs (32%, 57% and 77% in the charged materials). The covalence seems to stabilize higher oxidation states and thus enables a higher degree of delithiation during charge. However, the extent of irreversible reactions, especially on the surface of the particles, is also prone to increase at a higher degree of delithiation and at a related larger absence of ionic $Ni^{2+}$, which shortens the material lifetime.

The degree of covalence of the NCMs also determines the electrochemical potential and thus the electrode potential and cell voltage, **Figure 8**, *i.e.* apart from capacity, the second parameter important for the energy density of Li-ion batteries. **Figure 8A** shows the molecular orbital scheme of a transition metal in an octahedral O environment while **B** gives a schematic depiction of the bonding, non-bonding and antibonding states. The spontaneous reaction of layered oxides is the intercalation of Li-ions, which is the well-known discharge reaction. This implies that electrons are taken out of the highest occupied molecular orbital (HOMO) of the anode and are filled into the lowest unoccupied molecular orbital (LUMO) of the cathode (electron $e^-$ and blue arrows, **Figure 8C**). Therefore, the LUMOs of the NCMs define the redox potential of the cathodes and thus the open cell voltage of the batteries (**Figure 8C**).[35,36] The LUMOs of the NCMs, in turn, are antibonding *Me*-O* states. With an increasingly covalent character, the bonding states decrease while the antibonding states increase in energy, as depicted in **Figure 8** (left side). Consequently, the LUMO rises in energy and the mean discharge voltage decreases. Due to the disproportionately high increase of covalent $Ni^{3+}$ (**Figure 9A**) from NCM 111 (14%) *via* NCM 622 (37%) to NCM 811 (55%), the covalent character of the materials increases and thus the mean discharge voltage decreases (NCM 111: 3.84 V, NCM 622: 3.77 V, NCM 811: 3.72 V). In comparison: the $Ni^{2+}$ content in discharged NCM 111 is 19%, in NCM 622 23% and in NCM 811 25% (**Figure 9A**). Reducing the amount of covalent $Ni^{3+}$ in pristine NCMs would therefore lower the LUMOs, leading to higher mean discharge voltages. **Figure 9B** depicts semi-quantitatively the $Ni^{2+}$/capacity and the $Ni^{3+}$/voltage relationship as well as the energy density of the NCMs. The capacity increases almost linearly with the $Ni^{2+}$ content. With an increasing $Ni^{3+}$ content, the voltage decreases. The increasing capacity and the decreasing voltage affect the energy density in opposite directions. Therefore, the difference in energy density between NCM 811 and NCM 622 (671 Wh/kg – 669 Wh/kg = 2 Wh/kg) is neglectable. The difference between NCM 622 and NCM 111 (669 Wh/g - 615 Wh/kg = 54 Wh/kg) is larger due to the higher $Ni^{2+}$ content and thus the higher capacity of NCM 622 (NCM 622: 23% $Ni^{2+}$/178 mAh/g, NCM 111: 19% $Ni^{2+}$/160 mAh/g).

### 3. Conclusion

The present work reveals that the reversible charge compensation in NCMs proceeds from $Ni^{2+}$ to $Ni^{3+}$ while the electron density is shifted from O towards Ni forming covalent Ni-O bonds (96% $3d^8\ 2p^6$/> 4% $3d^9\ 2p^5 \leftrightharpoons$ 23% $Ni\ 3d^7\ O\ 2p^6$/77% $Ni\ 3d^8\ O\ 2p^5$). The high capacities of Ni-rich cathode materials are thus obtained due to the change in covalence, and not, as assumed before, due to cationic and/or anionic redox processes at individual *Me* (*Me* = Ni, Co) and O sites. The lack of $Ni^{2+}$, which is a very ionic Ni species compared to the more covalent $Ni^{3+}$, limits the reversible capacity of the materials. Simply increasing the Ni content of layered oxides without specific attention to the oxidation state of the introduced Ni, however, does hardly lead to higher capacities, as the $Ni^{2+}$ content stays fairly constant when going from NCM 111 (19%) to NCM 622 (23%) and finally NCM 811 (25%). Instead, the covalent character of the NCMs, more precisely the amount of covalent $Ni^{3+}$, increases from 14% (NCM 111) *via* 37% (NCM 622) to 55% (NCM 811), which lowers the mean discharge voltages.

We can learn from these findings, that the ability to increase the $Ni^{2+}$ content of Ni-rich layered oxides without increasing the amount of more covalent $Ni^{3+}$ will lead to outstanding capacities AND voltages. This opens potential areas of material design including anionic doping with mono- or divalent anions possessing a strong electron donor character and thus more $Ni^{2+}$. The results show, that the energy density of present lithium-ion batteries can still be increased significantly without the need of massive changes in cell chemistry.

## 4. Experimental Section

*Material preparation:* Details about the synthesis and characterization of the discussed cathode and reference materials are discussed in **S6** $LiNi_{1/3}Co_{1/3}Mn_{1/3}O_2$ (NCM 111), $LiNi_{0.6}Co_{0.2}Mn_{0.2}O_2$ (NCM 622) and $LiNi_{0.8}Co_{0.1}Mn_{0.1}O_2$ (NCM 811) were synthesized using a precipitation method.[52,53] $LiNiO_2$ (LNO) and $LiCoO_2$ (LCO) were synthesized using a sol-gel approach [54] and $Li_2MnO_3$ (LLO) was synthesized *via* a solid-state route.[55] Electrode preparation was performed using 92.5 wt-% of the cathode material (powder), 4 wt-% Imerys Super C65, 3.5 wt-% PvDF (Solef, Polyvinylidene difluoride, PvdF) and NMP (N-methylpyrrolidone) as processing solvent (6.5 mg). The ink was casted on an Al-foil (250 µm wet film thickness) and dried at 80°C. 11 mm in diameter samples were punched out of the dried electrodes and were assembled to a battery cell in an argon filled glovebox using Swagelok cells with two glass-fiber separators (11 mm in diameter, glass micro fiber 691, VWR, Germany), 60 µL LP57 (1 M $LiPF_6$ in ethylene carbonate (EC):ethyl methyl carbonate (EMC), 3:7 by weight, <20 ppm of $H_2O$, BASF SE, Germany) and lithium foil as opposite electrodes (11 mm in diameter, 99.9% Rockwood Lithium, United States). The underlying capacity used to calculate the C-rate for NCM 111, NCM 622, NCM 811, LNO, LCO and LLO was 160 mAh/g, 180 mAh/g, 180 mAh/g, 180 mAh/g, 140 mAh/g, 250 mAh/g, respectively. The applied voltage cut-offs upon galvanostatic cycling were chosen to be 4.3 V and 3.0 V.

*Sample Preparation:* For NEXAFS sample preparation different electrodes were cycled with C/20 at 25°C to the charged state, to a remaining capacity of 150 mAh/g, 100 mAh/g and 50 mAh/g in the first discharge and to the discharged state. After cycling, the cells were disassembled in an argon filled glovebox. Additional NCM 111 samples cycled for 100 cycles were rinsed with dimethyl carbonate (DMC, BASF SE, Germany) after disassembly in order to probe the impact of electrolyte decomposition products on the fluorescence yield soft XAS spectra (see **S7**). The obtained samples were dried in a Büchi oven ($10^{-3}$ mbar, 25 °C) prior to the measurements without contact to air in order to avoid outgassing in the UHV chamber. An argon filled transfer case was used for the sample transfer from the glovebox to the load lock of the beamline since NCMs (especially in their charged state) are air sensitive.

*NEXAFS measurements:* NEXAFS measurements were performed at IQMT's soft X-ray beamline WERA at the Karlsruhe synchrotron light source KARA (Germany). NEXAFS measurements at the Ni $L_{2,3}$, Co $L_{2,3}$, Mn $L_{2,3}$, O $K$ and F $K$ edge were carried out in fluorescence yield (FY) detection mode. For the Co $L_{2,3}$ edge measurements, the FY detection window (width 0.24 keV) was shifted up in energy from its nominal centre position in order to minimize crosstalk of the signal from the preceding F $K$ edge (see discussion in **S7**). In the case of the Mn $L_{2,3}$ edge, inverse partial FY detection (IPFY) [55] using the O $K$ fluorescence was employed. The photon-energy resolution in the spectra was set to 0.2-0.4 eV. Energy calibration (using a NiO reference), dark current subtraction, division by $I_0$, background subtraction, data normalization and absorption correction was performed as described in [56,57]. The background treatment of the Ni and Co $L_{2,3}$ edges was more challenging due to the presence of F K EXAFS features and a detailed description is given in **S7**. Data evaluation of the *Me* $L_{2,3}$ edges was done using charge-transfer multiplet calculations (CTM4XAS, Crispy and Quanty).[17–20] The calculated spectra were fitted to the measured data with an Excel macro written by one of us, applying a least square method to minimize the difference in area between the data and the fit by shifting the calculated spectra along the energy axis and varying the intensity. The O K edge peaks were fitted with three Gaussian functions (intensity and position of peaks were kept variable) using the program Fityk as shown in **S5**.

*Powder diffraction:* Time-resolved in situ synchrotron X-ray powder diffraction (SXPD) of the first discharge was performed at beamline I11 (Diamond Light Source, UK) using the position sensitive detector (PSD) of the beamline.[58] The energy of the x-ray beam was tuned to 25 keV and the calibrated wavelength was 0.489951(10) Å. The battery cells were mounted on a xyz-stage and each cell was adjusted to the centre of diffraction. The 2D data was refined using the software package Fullprof (2θ range: 0 - 65°).[59] Due to preferential orientations of Al (current collector of the NCM electrode) and Li (opposite electrode), the phases were included in the refinements using the Le Bail method. The structural input parameters for the refinements of NCM 111, NCM 622 and NCM 811 (in the pristine state) were obtained from crystallographic data files.[60–62] Upon charge anisotropic microstrain is increasing, which was implemented as described in [13].

## 5. Supporting Information

Supporting Information is available from the Wiley Online Library or from the author. It includes Mn $L_{2,3}$ and Co $L_{2,3}$ NEXAFS spectra of charged and discharged NCM 111, NCM 622 and NCM 811, a discussion about the choice of reference materials, the theoretical background of charge-transfer multiplet calculations and additional information how Me $L_{2,3}$ edge spectra are simulated, fits of O K edge spectra, details about the synthesis and characterisation of the cathode and reference materials as well as a discussion about the background observed in the NEXAFS spectra and the best way of sample preparation.

## 6. Acknowledgement

Karin Kleiner and Michael Merz contributed equally to this work. We wish to acknowledge KARA and the KNMF (both Karlsruhe, Germany), as well as the Diamond Light Source (UK, proposal NR19772) for provision of beamtime. We also acknowledge Christoph Suergers for the provision of the transfer chamber which enabled us to transfer our samples from the Glovebox into the load lock of the WERA beamline at KARA without any contact to air. Financial support was provided by the Federal Ministry of Education and Research (BMBF) under the funding number 03XP0231.

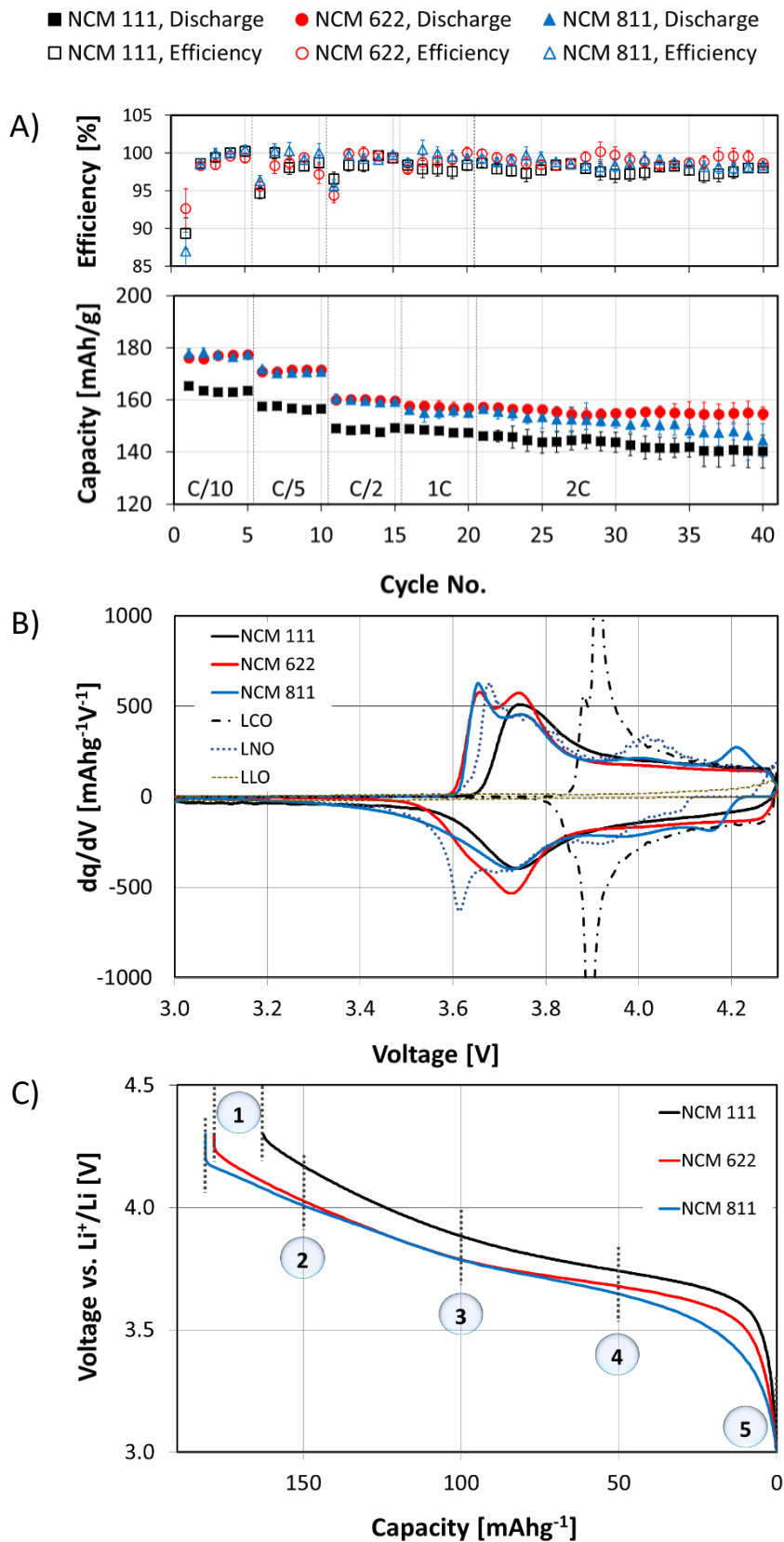

**Figure 1:** (A) Coulomb efficiency (upper panel) and discharge capacity (lower panel) of NCM 111 (black squares), NCM 622 (red circles) and NCM 811 (blue triangles), obtained from 3 x

40 cycles for each material at different C-rates. (B) Differential capacity (dq/dV) plot of the first NCM cycles as well as of the reference materials LNO (LiNiO$_2$), LLO (Li[Li$_{1/3}$Mn$_{2/3}$O$_2$) and LCO (LiCoO$_2$). (C) shows the first discharge cycle of the NCMs and the points 1-5, at which NEXAFS data was taken.

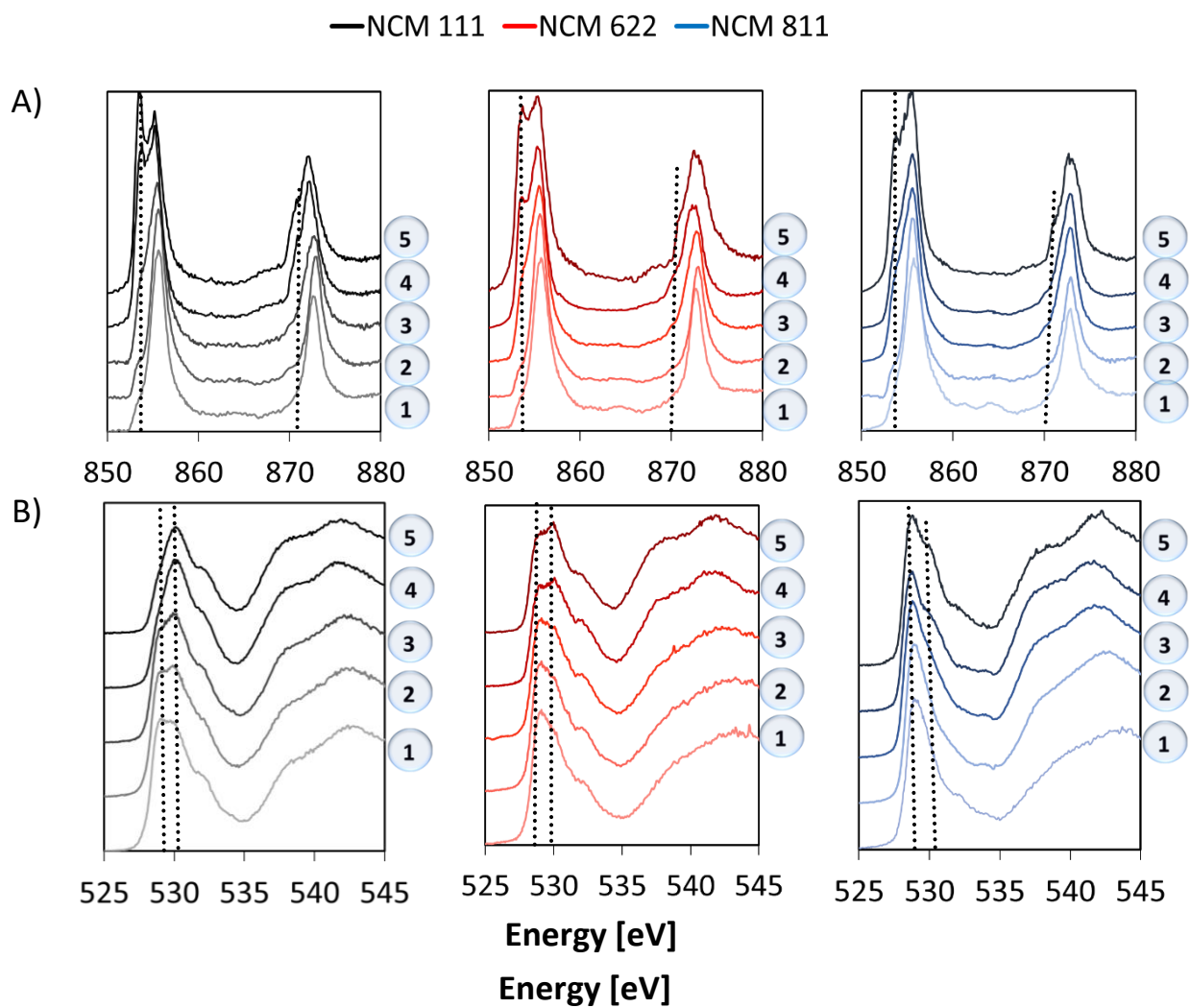

**Figure 2:** Ni $L_{2,3}$ (A) and O K edge XAS spectra (B) of charged (1), 150% state of charge (2), 100% state of charge (3), 50% state of charge (4) and discharged (5) NCM 111 (black), NCM 622 (red) and NCM 811 (blue), measured in fluorescence yield. Main changes are marked with dashed lines in the spectra.

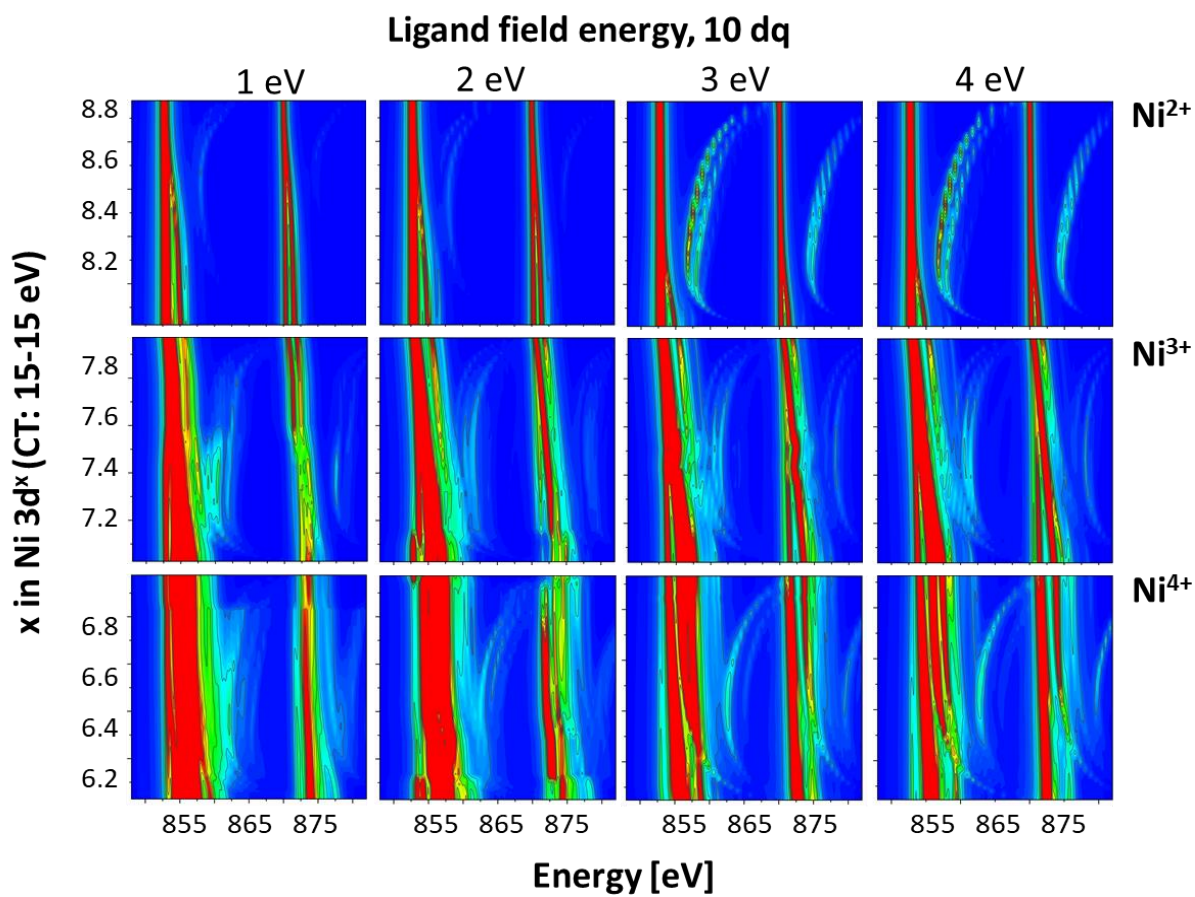

**Figure 3:** CTM calculations obtained for $Ni^{2+}$, $Ni^{3+}$ and $Ni^{4+}$ varying the ligand-field energy from 1 eV to 4 eV and the charge-transfer-energy from -15 eV to 15 eV. 'b' in Ni $3d^{a+b}$ O $2p^{6-b}$ (with x = a + b, horizontal axis) was obtained varying the charge-transfer-energy while 'a' is determined by the oxidation state of Ni ($Ni^{2+}$: a = 8, $Ni^{3+}$: a = 7, $Ni^{4+}$: a = 6).

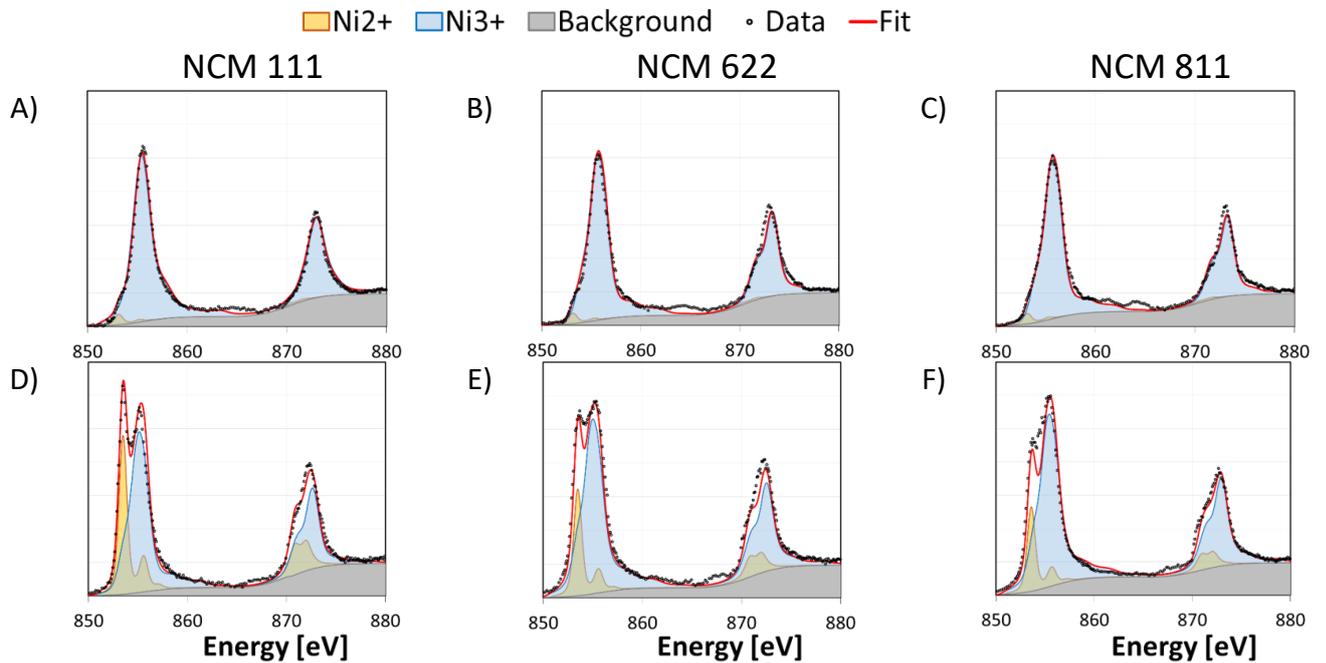

**Figure 4:** Measured Ni $L_{2,3}$ edge spectra (black dots) of charged (A-C) and discharged (D-E) NCM 111 (A, D), NCM 622 (B, E) and NCM 811 (C, F), simulated with a covalent $Ni^{3+}$ configuration (23% $3d^7\,2p^6 \rightleftharpoons$ 77% $3d^8\,2p^5$, blue area) and an ionic $Ni^{2+}$ configuration (< 96% $3d^8\,2p^6 \rightleftharpoons$ > 4% $3d^9\,2p^5$, yellow area). The grey area is the background and the red line the superposition of the simulation.

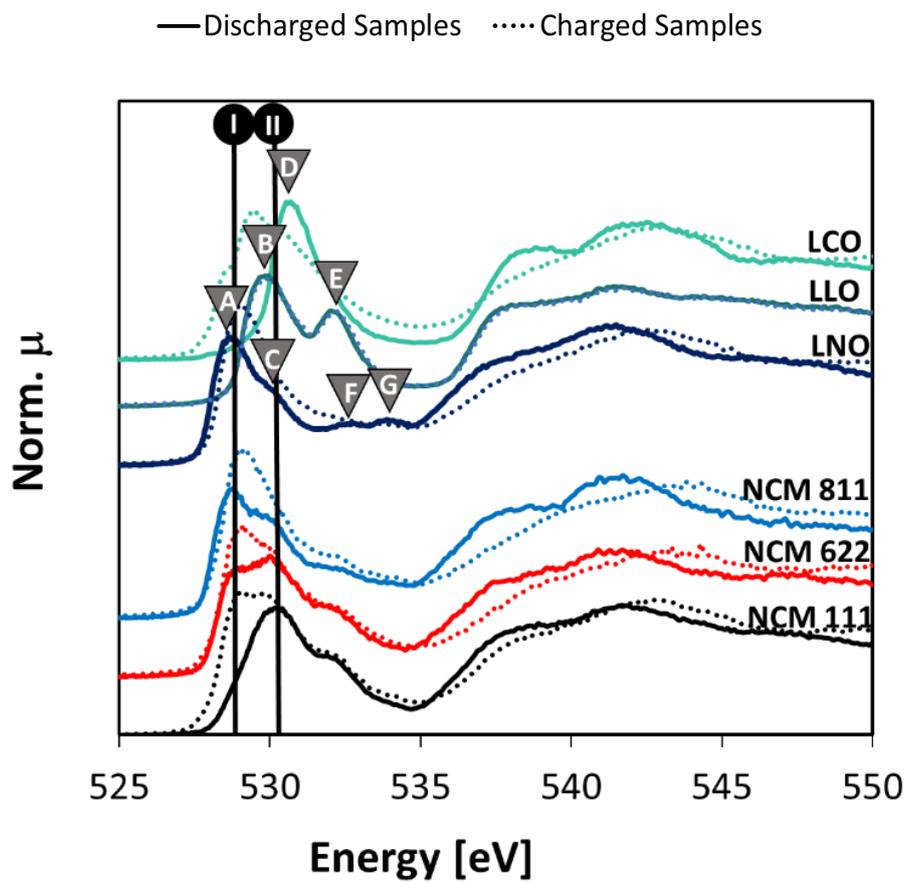

**Figure 5:** O K XAS spectra of charged (dotted lines) and discharged (solid lines) NCM 111, NCM 622, NCM 811, LNO (LiNiO$_2$), LLO (Li[Li$_{1/3}$Mn$_{2/3}$O$_2$]) and LCO (LiCoO$_2$).

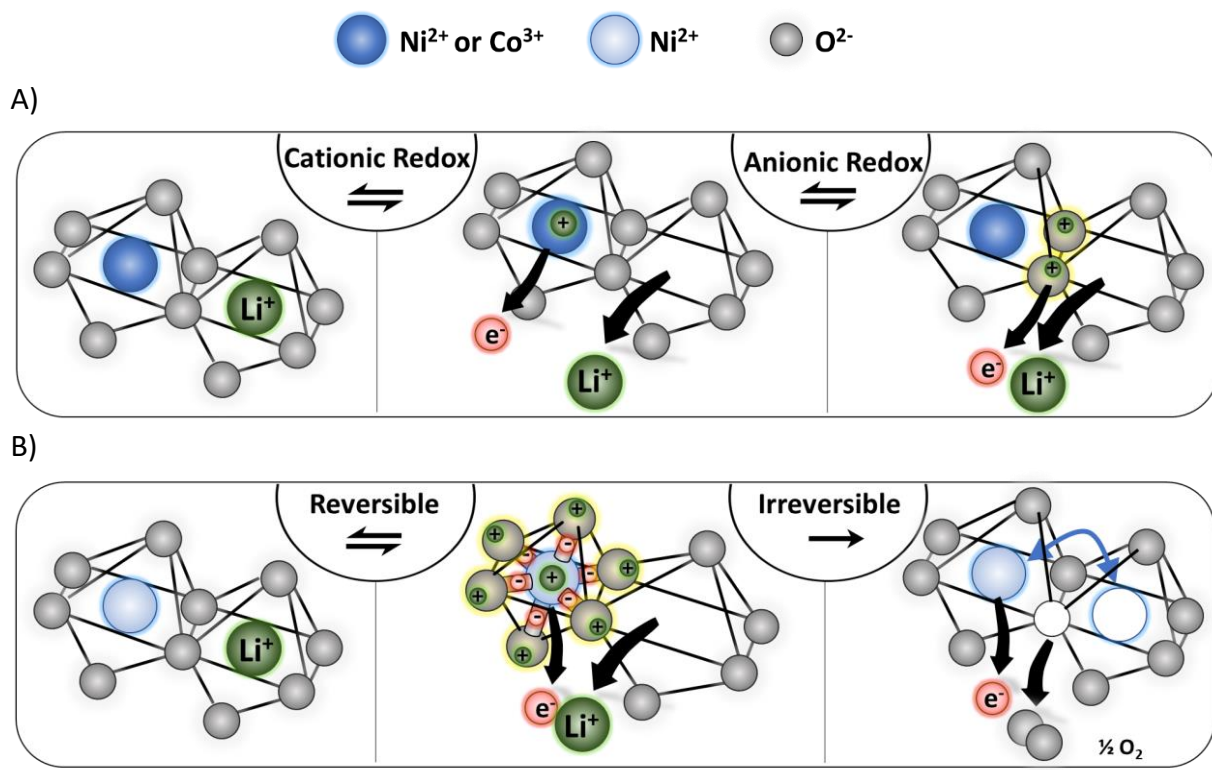

**Figure 6:** (A) Schematic depiction of the reversible cationic (middle panel) and anionic (right panel) redox process in layered oxide as discussed in literature. (B) The first and second panel show the reversible oxidation of $Ni^{2+}$ to $Ni^{3+}$ upon charge while electron density is shifted from the O ligands towards Ni (hybridization of Ni-O bonds). By the point, where the $Ni^{2+}$ content is suppressed to zero, irreversible oxygen release and phase transformations set it (last panel).

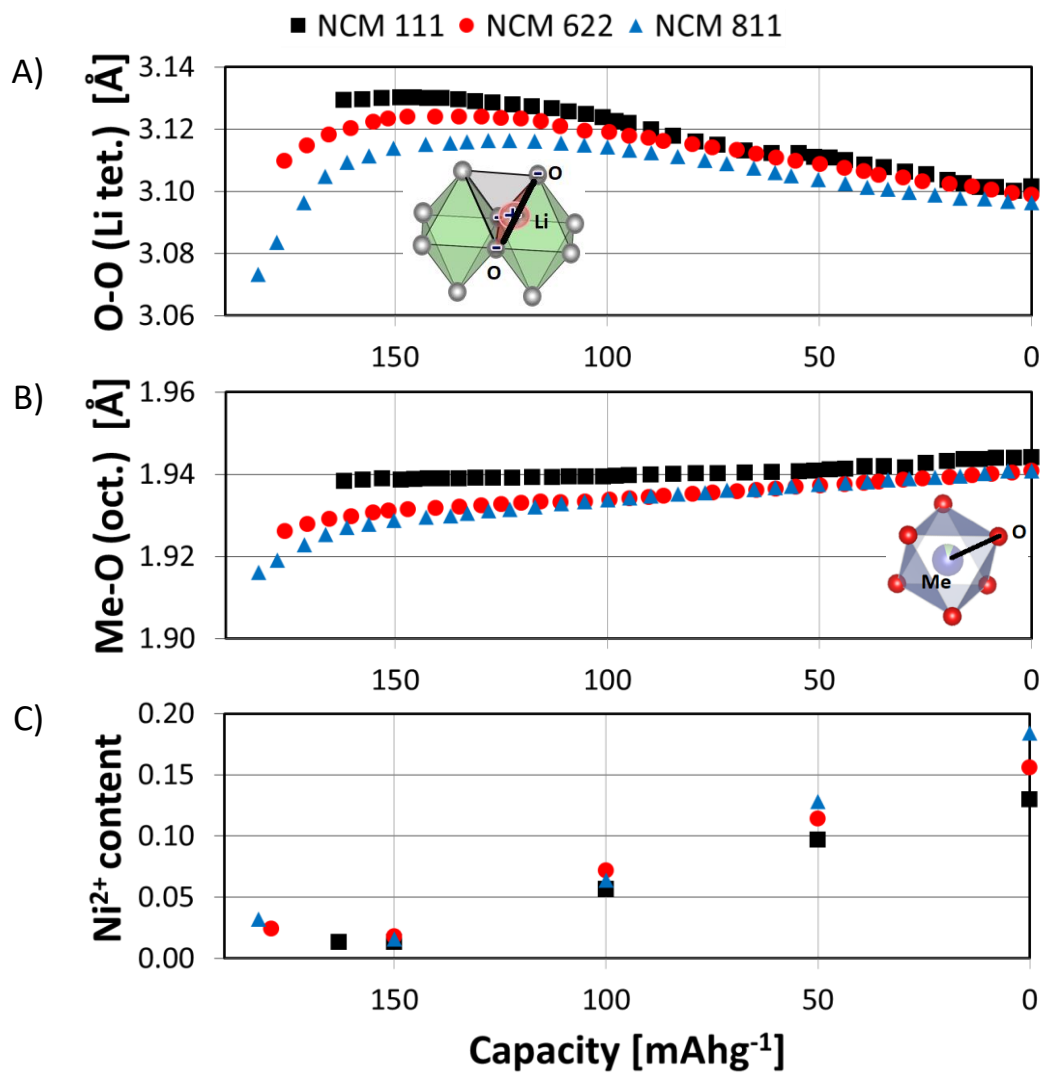

**Figure 7:** Results obtained upon the first discharge of NCM 111, NCM 622 and NCM 811 from *operando* Synchrotron X-ray powder diffraction (A, B) and near edge X-ray absorption finestructure spectroscopy (C). A) gives the O-O distance of the tetrahedral sites in the Li-layer of the rhombohedral NCM structure while B) shows the evolution of the *Me*-O bond length. In (C) the $Ni^{2+}$ stoichiometry, obtained from simulations of the Ni $L_{2,3}$ NEXAFS spectra, is depicted.

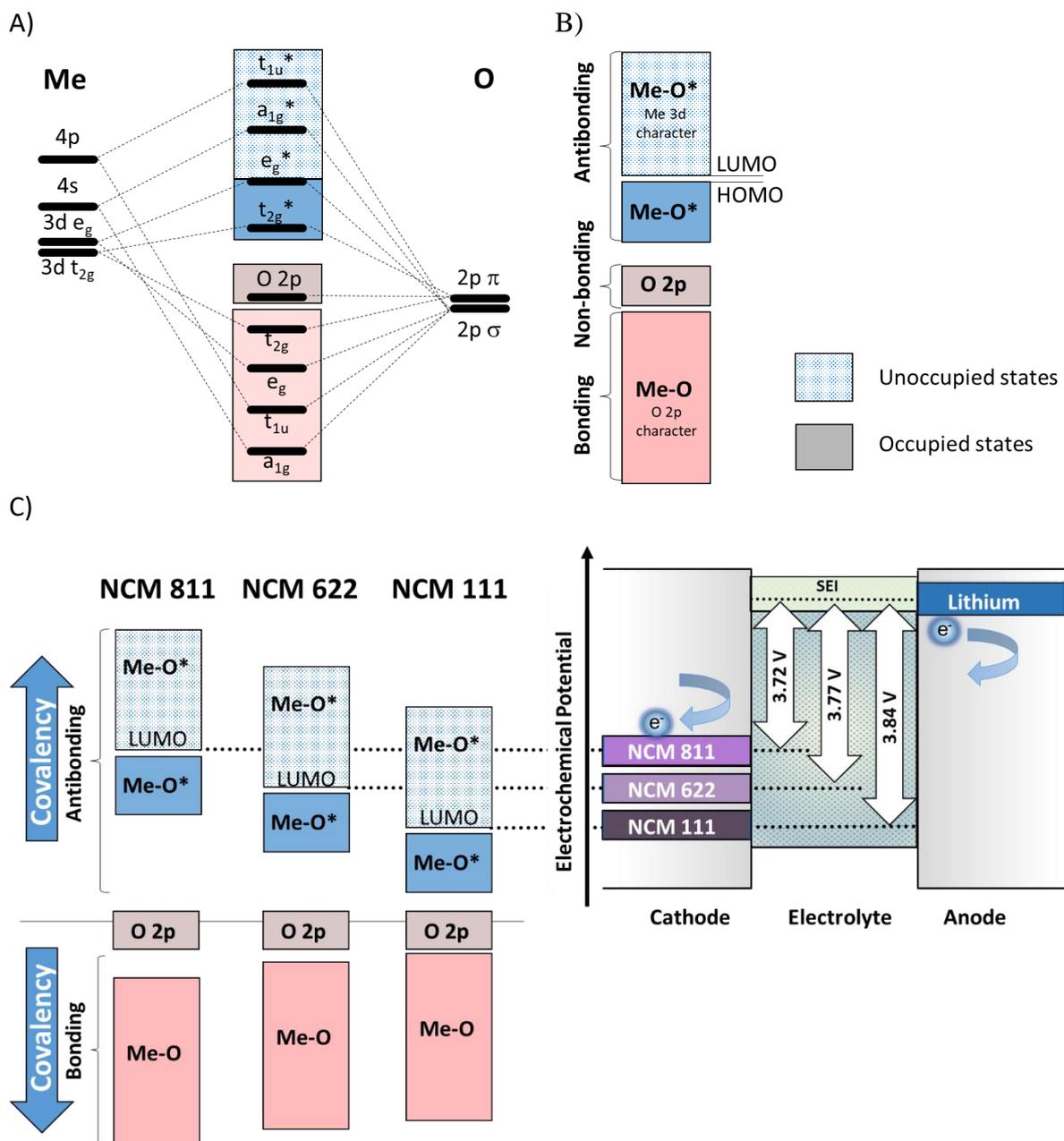

**Figure 8:** (A) Molecular Orbital Scheme of a transition metal (Me) in an octahedral oxygen (O) environment and (B), a schematic depiction of bonding, non-bonding and antibonding states. (C) With and increasing covalent character from NCM 111 *via* NCM 622 to NCM 811, bonding orbitals decrease and antibonding orbitals increase in energy (left side). This shifts the LUMO up in energy and thus, the cell voltage decreases (right side).

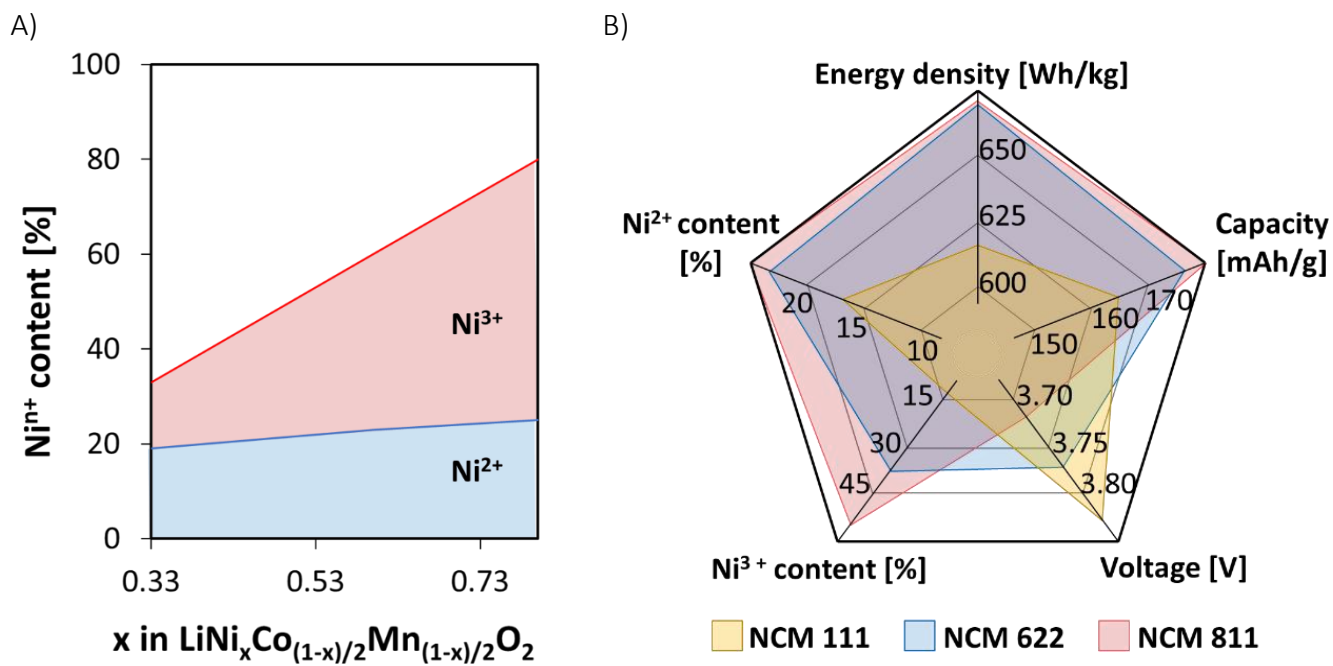

**Figure 9:** (A) The $Ni^{2+}$ and $Ni^{3+}$ content in dependence on the overall Ni-content and (B), the energy density, capacity, voltage, $Ni^{2+}$ and $Ni^{3+}$ relation in NCM 111, NCM 622 and NCM 811.

**Table of contents**

Unraveling the redox active centers in present cathode materials, showing how the exchange of electrons and ions is compensated upon charge and discharge and correlating the results with crystallographic information has revealed, why the energy density of Li-ion batteries is limited and how these limitations can be overcome in future.

**Keyword:** Energy density limitation in NCMs

*Karin Kleiner\*, Claire A. Murray, Cristina Grosu, Sarah J. Day, Martin Winter, Peter Nagel, Stefan Schuppler, Michael Merz*

**On the microscopic origin of reversible and irreversible reactions of LiNixCoyMnx cathode materials: Ni-O hybrid bond formation vs. cationic and anionic redox**

**ToC figure**

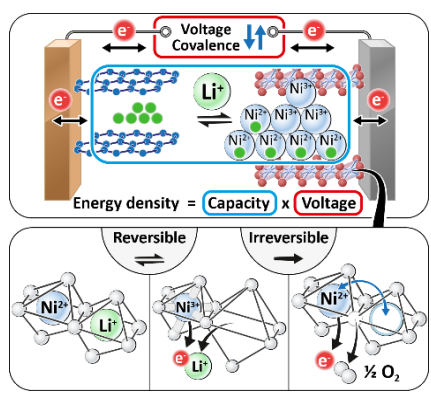

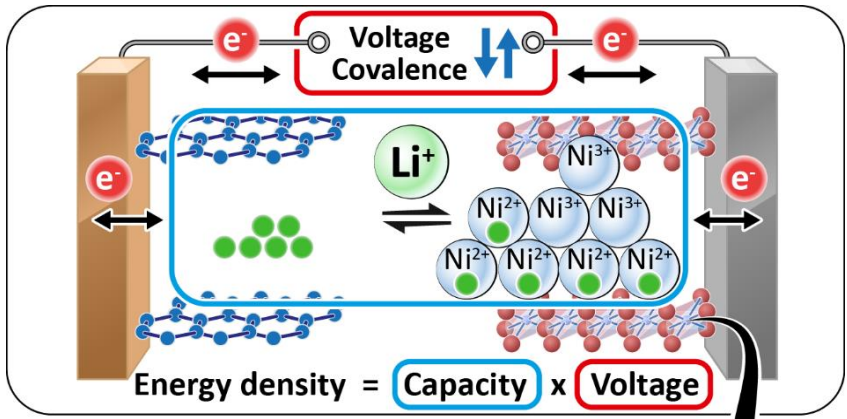

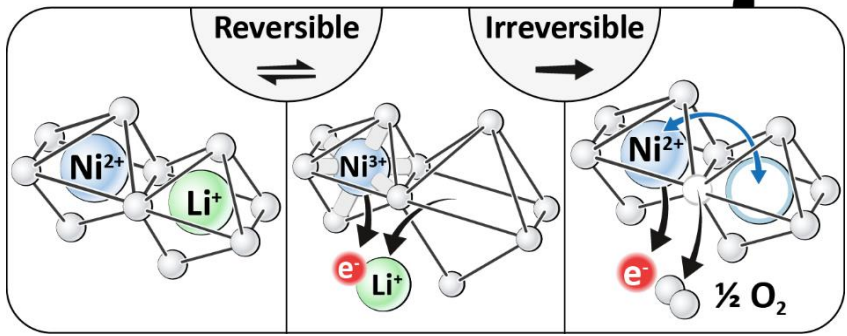

# Supporting Information

**How to improve Li-ion batteries: Microscopic origin of their limitations**

*Karin Kleiner\*, Claire A. Murray, Cristina Grosu, Sarah J. Day, Martin Winter, Peter Nagel,*

*Stefan Schuppler, Michael Merz*

Dr. Karin Kleiner, Prof. Martin Winter
Münster Electrochemical Energy Technology, University of Muenster (MEET, WWU)
Corrensstr. 46,
48149 Muenster, Germany
E-Mail: karin.kleiner@wwu.de

Dr. Claire Murray, Dr. Sarah Day
Beamline I11, Diamond Light Source (I11, DLS)
Harwell Science and Innovation Campus, Diamond House
Didcot OX11 0DE

Cristina Grosu
Institute of Energy and Climate Research, Research Center Jülich (IEK9, FZJ)
52425 Juelich, Germany

Prof. Martin Winter
Helmholtz-Institute Münster, Forschungszentrum Jülich GmbH
Corrensstr. 46,
48149 Muenster, Germany

Dr. Peter Nagel, Dr. Stefan Schuppler, Dr. Michael Merz
Institute for Quantum Materials and Technologies, Karlsruhe Institute of Technology (IQMT, KIT)
Hermann-von-Helmholtz-Platz 1
76344 Eggenstein-Leopoldshafen, Germany

## S1. Choice of reference materials

As reference materials rhombohedral LCO (LiCoO$_2$, Co$^{3+}$ Low spin) and LNO (LiNiO$_2$, Ni$^{2+/3+}$ low spin) were used. Mn in LiMnO$_2$ has a t$_{2g}^4$ e$_g^1$ (formal Mn$^{3+}$) configuration and is therefore not appropriate for a comparison to a t$_{2g}^3$ e$_g^0$ (formal Mn$^{4+}$). Instead, LLO (Li$_2$MnO$_3$ = Li(Li$_{1/3}$Mn$_{2/3}$O$_2$), which is structurally related to rhombohedral layered oxides, served as reference.[1] In order to ensure the comparability of the reference spectra to the data obtained from the NCMs, the Co $L_{2,3}$, Ni $L_{2,3}$ and Mn $L_{2,3}$ spectra of charged and discharged LNO, LCO and LLO are shown in **Figure S1**. All discharged spectra look similar to the corresponding Ni, Co and Mn spectra of NCM 111, NCM 622 and NCM 811. The charged Co L and Mn L spectra of LCO and LLO show slight changes if compared to the discharged ones. While in NCM neither Co nor Mn is electrochemically active, the transition metals show a certain redox activity if no Ni is present.

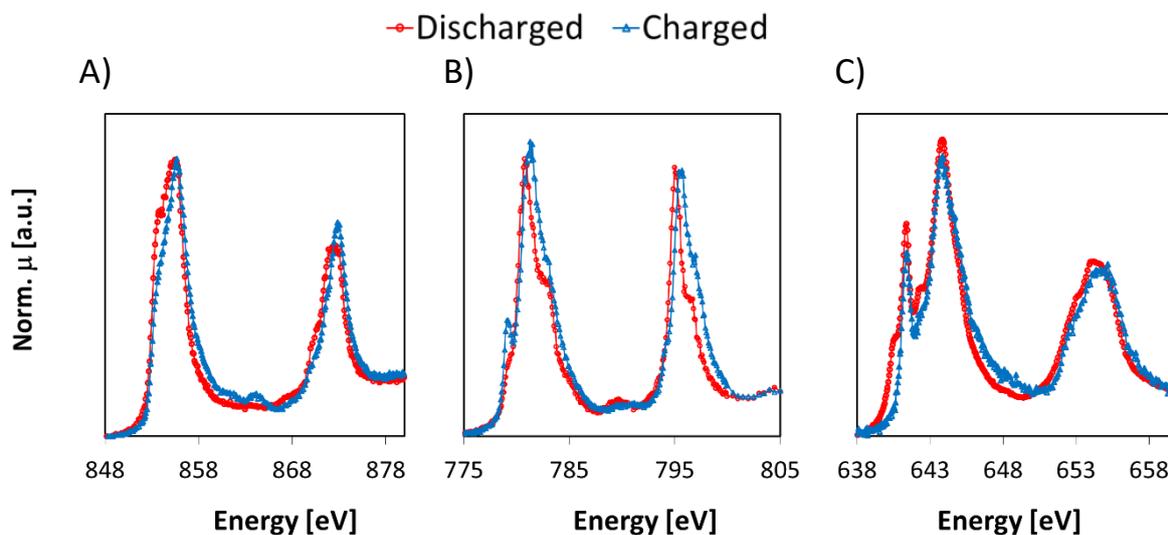

**Figure S1:** (A) Ni $L_{2,3}$ edge of LNO, (B) Co $L_{2,3}$ edge of LCO and (C), Mn $L_{2,3}$ edge of LLO, measured in the charged (red) and in discharged state of the reference materials.

## S2. Co L and Mn L edge of NCM 111, NCM 622 and NCM 811

For completeness, the Co $L_{2,3}$ (A) and Mn $L_{2,3}$ of charged and discharged NCM 111, NCM 622 and NCM 811 are shown in **Figure S2**. Upon the first discharge no changes are observed showing that Co and Mn are not redox active in the bulk material. This is in contrast to literature which is discussed in the main text.

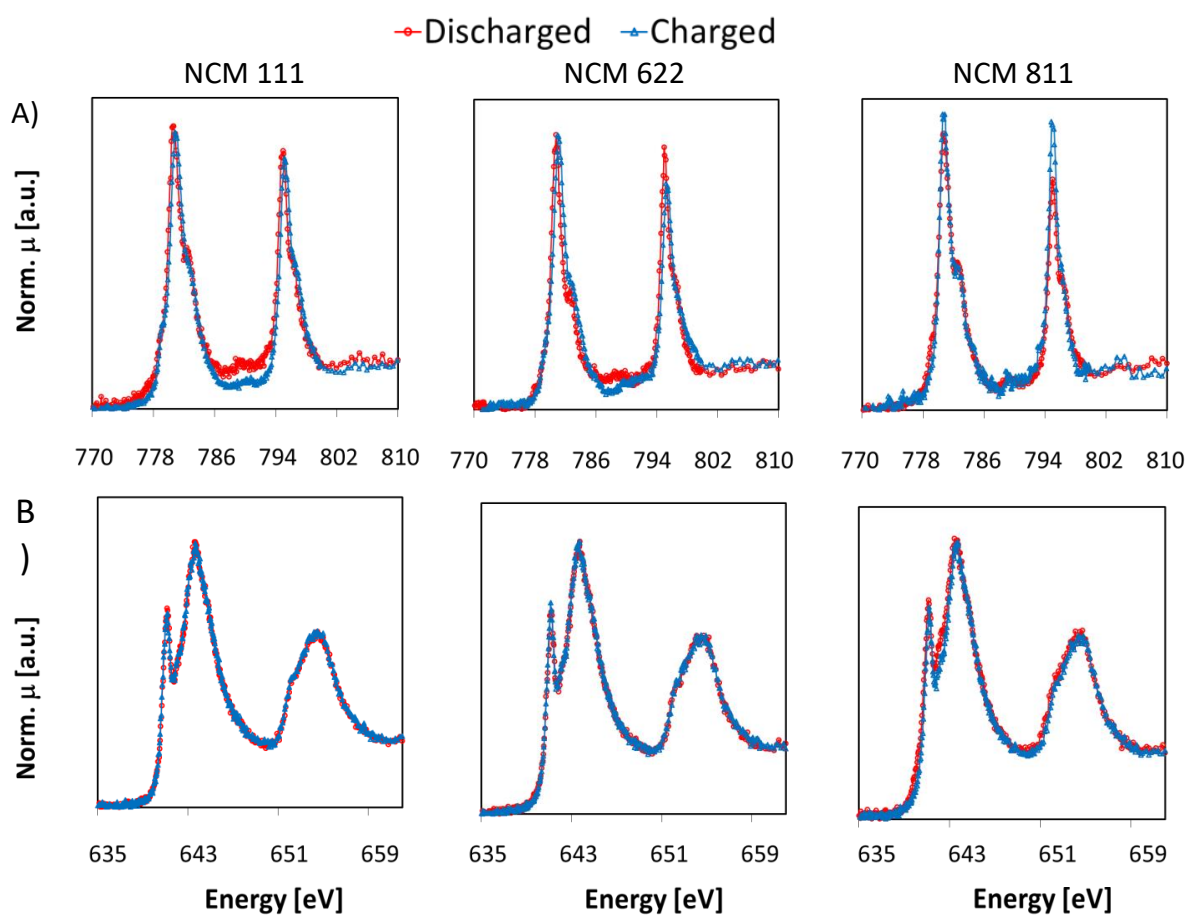

**Figure S2:** Co $L_{2,3}$ (A) and Mn $L_{2,3}$ edge XAS spectra (B) of charged (blue) and discharged (red) NCM 111, NCM 622 and NCM 811.

## S3. Charge-transfer Multiplet Theory

Density functional theory (DFT) is the most successful approach to describe the ground state of layered oxides, often depicted with the density of states (DOS), **Figure S3**. These calculations are used to describe the band structure along the Brillouin zone (local-density approximation, LDA) of an infinite crystal using the linear augmented plane wave method.[2] From the crystal potential of these DFT calculations a set of Wannier orbitals, with a sufficient amount of ligand orbitals which allows to restrict the calculations to one $MeO_6$ octahedron, are taken. Thereby the band structure reduces to O p and Ni d-like $\pi/\pi*$ and $\sigma/\sigma*$ levels. Although these LDA calculations predict layered oxides to be metallic and the materials are found to be semiconductors or insulators, the set of Ni 3d and O 2p orbitals span together the low-energy solution of the Schroedinger equation for the crystal potential sufficiently and therefore constitute a good single-particle basis set for the many body calculations. Charge fluctuations or their suppression (depicted in **Figure S3A**) further alter the DOS. The energy needed for such electron exchanges (or their suppression) triggers the band gap and therefore defines the type of conductivity. Charge fluctuations between Me 3d-states ($d_i^n + d_j^n \rightarrow d_i^{n-1} + d_j^{n+1}$), **Figure S3A** (1) can be suppressed due to large Coulomb interactions $U_{dd}$ between the d electrons (up to 10 eV for late transition metals), introducing a band gap between occupied and unoccupied states, (**Figure S3B** and **C**).[3] The HOMOs and LUMOs (highest and lowest occupied/unoccupied molecular orbitals) split up into a lower and a higher Hubbard band and the materials (formerly considered metallic, **Figure S3B**) become Mott Hubbard semiconductors or insulators (**Figure S3C**). The band gap also depends on the electronegativity of the ligands indicating the presence of charge fluctuations from the ligands towards the metals ($d^n \rightarrow d^{n-1} L$, where L denotes an electron hole in the 2p states of the ligands, **Figure S3A** (2).[4] If the ligand to metal charge-transfer (LMCT) energy $\Delta$ is lower than $U_{dd}$, strong interactions between Me 3d and O 2p orbitals can occur (**Figure S3D**) and the *Me*-O bonds become highly covalent in nature. The excited state in the X-ray absorption process ($2p^6 3d^n + h\nu \rightarrow 2p^5 3d^{n+1}$) contains, in addition to the ground state, 2p spin-orbit couplings and 2p-3d interactions of the core hole and electrons in d-orbitals (multiplet effects) included as well in the many body approach of the CTM calculations. The effects discussed above are implemented in the CTM calculations using an Anderson impurity model.[2,5]

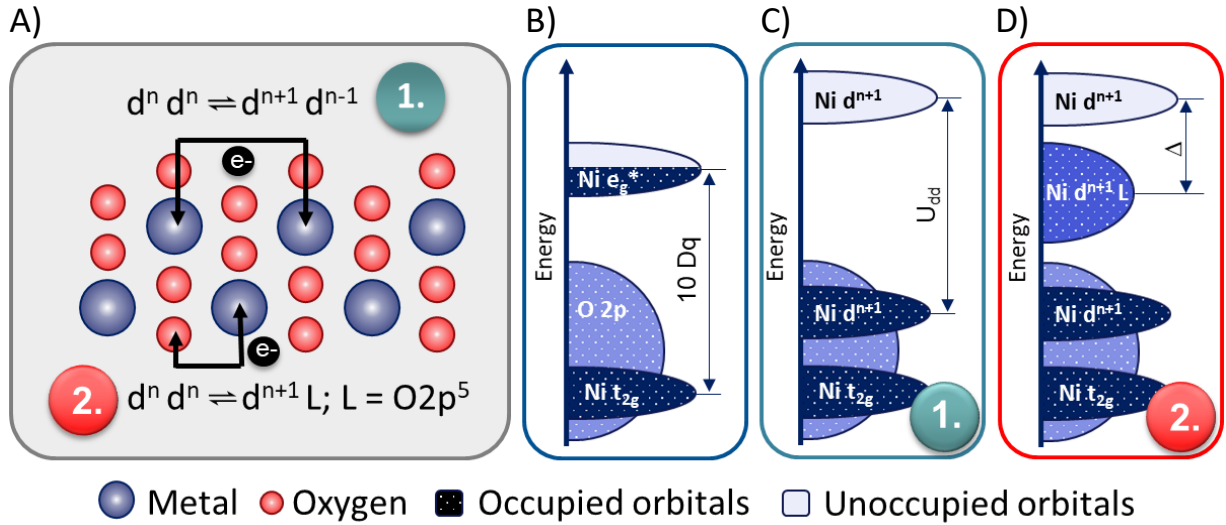

**Figure S3:** Depiction of charge fluctuations in layered oxides (A), the DOS without charge fluctuations (B), the DOS with the assumption of repulsive coulomb interactions between d-electrons $U_{dd}$ (C) and with a ligand metal charge-transfer LMCT $\Delta$ (D).

The above discussed charge fluctuations and their suppression as well as Coulomb interactions alter the X-ray absorption process, leading to satellites and to a contraction of the multiplet structure in the XAS spectrum. In the present work, XAS spectra are calculated using CTM4XAS, Quanty and/or Crispy, which work based on the atomic multiplet theory described above.[6,7] The core hole potential $U_{pd}$ describing the interactions between the core hole and the additional electrons in Me 3d states ($2p^6\ 3d^{n+1}\ L + h\nu \rightarrow 2p^5\ 3d^{n+2}\ L$), the LMCT charge-transfer-energy $\Delta$, the coulomb interactions $U_{dd}$ ($E_{LMCT} = \Delta + U_{dd} - U_{pd}$) and the ligand-field energy 10 Dq (see **Figure S4**) are adjustable parameters in these calculations.

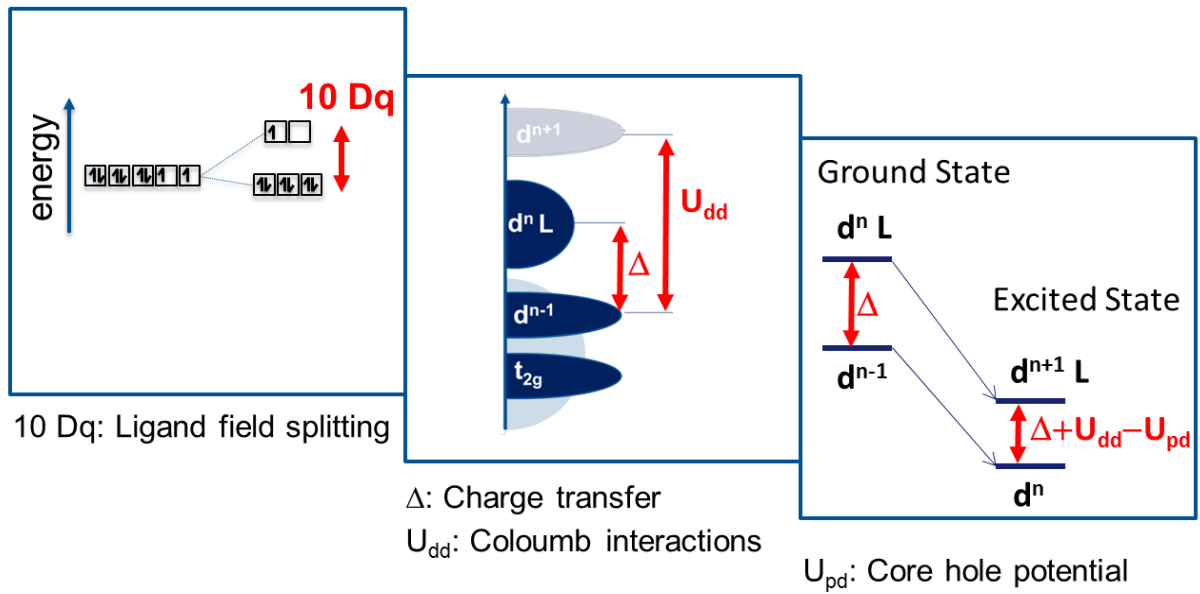

**Figure S4:** Variables in CTM calculations considering the Ligand-field 10 Dq, the core hole potential $U_{pd}$, the charge-transfer $\Delta$ and the coulomb interactions $U_{dd}$.

## S4. Fits of the Transition Metal L edges

Pristine and discharged Ni $L_{2,3}$ edges can be fitted with a superposition of a $Ni^{2+}$ (90% $3d^8$ and 10% $3d^9L$) and a $Ni^{3+}$ configuration (20-30% $3d^7$ and 70-80% $3d^8L$) as shown by fitting the Ni $L_{2,3}$ edge of NCM 111, NCM 622, NCM 811 in the charged state **A**, at a remaining capacity of 150 mAh/g **B**, 100 mAh/g **C**, 50 mAh/g (1$^{st}$ discharge) **D** and in the discharged state **E**, **Figure S5**. The electronic configurations of Ni in NCM 111, NCM 622 and NCM 811, calculated with CTM and the LF energy (10 Dq), the core hole potential minus the $d^n d^n$ Coulomb interactions ($U_{pd}$-$U_{dd}$), the LMCT ($\Delta$) and the observed line broadening are given in **Table S1**.

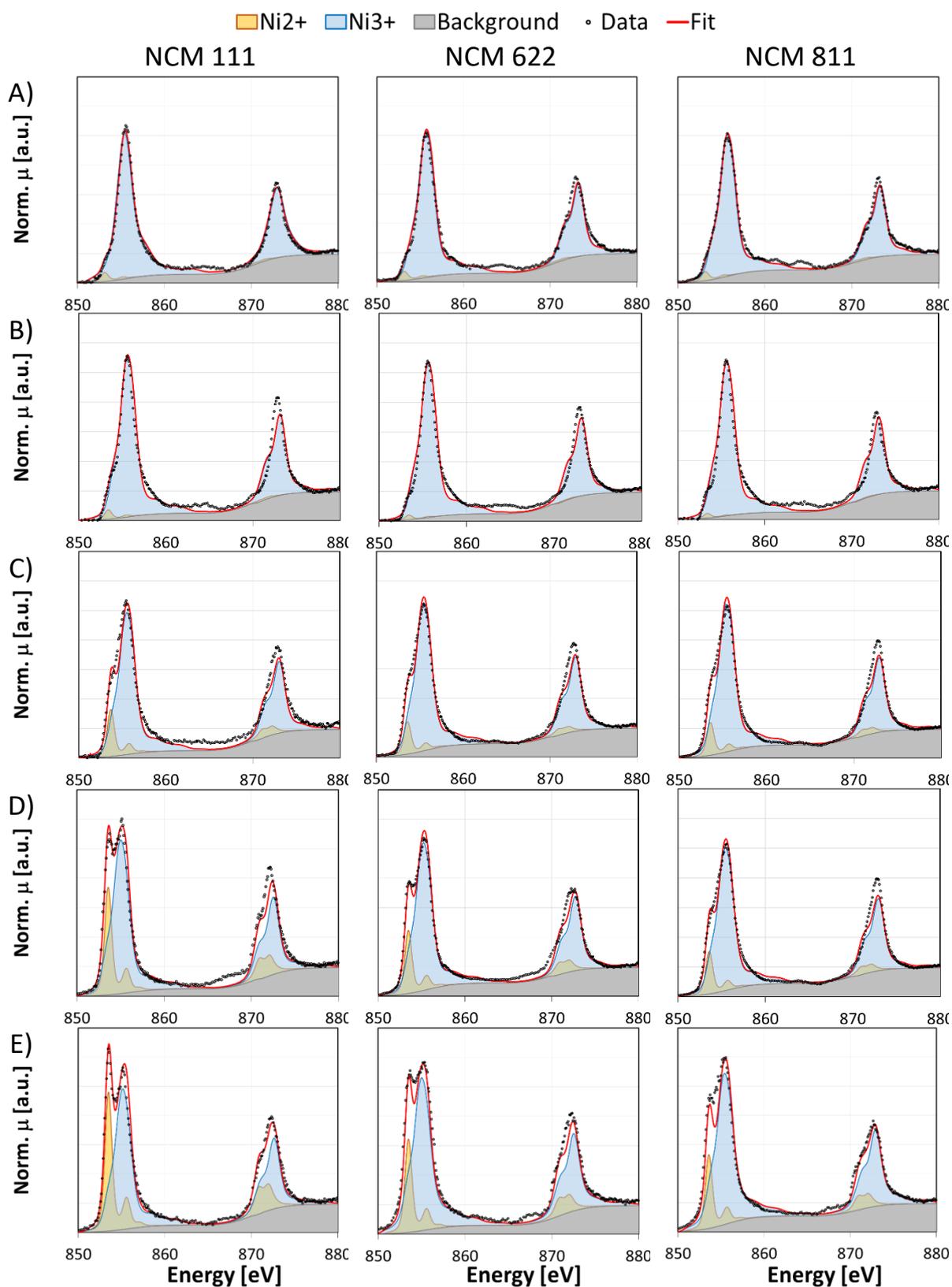

**Figure S5:** Ni $L_{2,3}$ edge of NCM 111, NCM 622 and NCM 811 charged to 4.3 V A) state, at 150 mAh/g B), 100 mAh/g C), 50 mAh/g D) and in the discharged state E), fitted with charge-transfer multiplet calculations.

**Table S1:** Ligand-field energy 10 Dq, Core hole potential $U_{pd}$ minus coulomb interactions $U_{dd}$, charge-transfer-energy $\Delta$ and Lorentzian and Gaussian broadening obtained from the CTM calculations of the Ni $L_{2,3}$ and edge. $Ni^3$ is considered to be a low spin (LS) configurations.

| | 10 Dq [eV] | $U_{pd}$ - $U_{dd}$ [eV] | $\Delta$ [eV] | Ltz/Gss (broadening) |
|---|---|---|---|---|
| $Ni^{2+}$, $t_{2g}^6 e_g^2$ | 1.5 | 1 | 8.0 (< 10% $d^9L$) | 0.3/0.3 |
| $Ni^{3+}$, $t_{2g}^6 e_g^1$ | 3.0 | 1 | -4.0 (77% $3d^8L$) | 0.5/0.3 |

**Figure S6A** shows the calculated $Mn^{4+}$ spectra, obtained by taking various, higher excited configurations such as $3d^{3+x}$ $L^x$ with $0 \leq x \leq 2$ into account.[8] The Mn $L_{2,3}$ edges were measured in inverse partial FY (IPFY) detection mode and therefore absorption corrections do not need to be performed. The Co $L_{2,3}$ edge is sitting on top of the EXAFS features of the F K edge (see **Figure S12**), which makes the normalization and interpretation of the spectra more difficult. Nevertheless, the spectra (although noisier compared to Ni) can be described with a $Co^{3+}$ low spin configuration (LF = 3.0 eV) as shown in (**Figure S6B**). LMCT interactions are minor effects since the spectra can be sufficiently described without assuming charge fluctuations from O to Co. An overview of the Co spectra used to describe the measured data is provided in **Figure S7**.

---

8    Merz et al., to be published.

A)

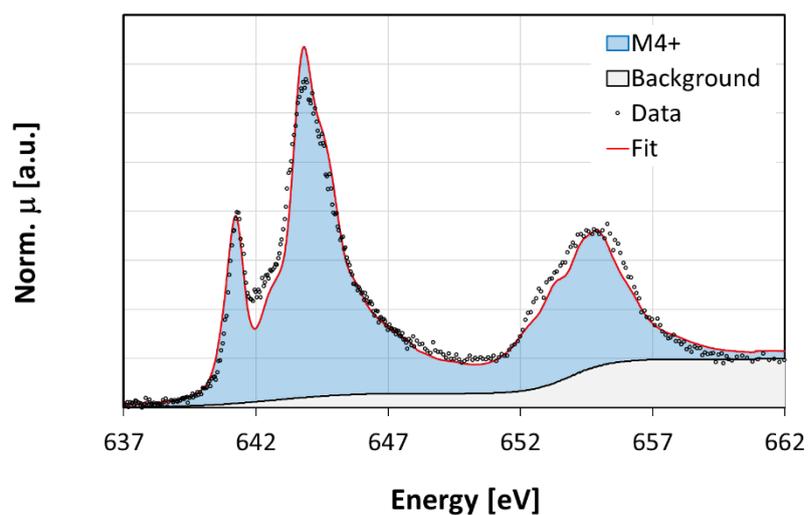

B)

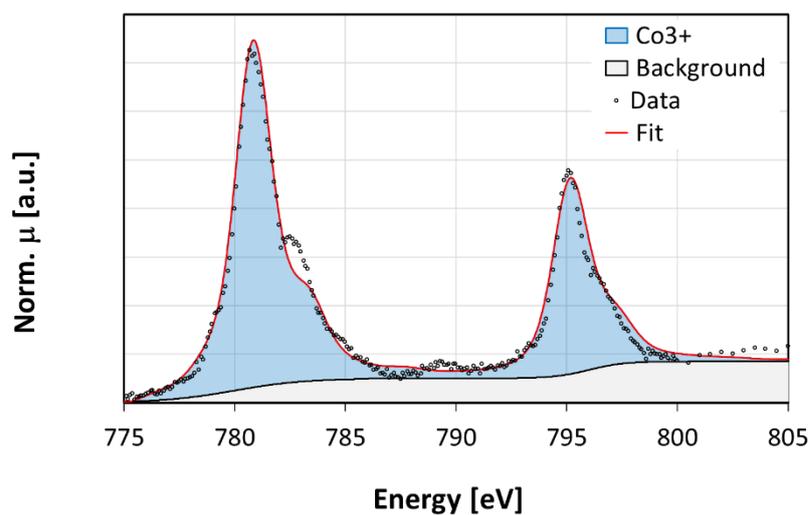

**Figure S6:** (A) Mn $L_{2,3}$ XAS spectra, calculated with Quanty. The Mn $L_{2,3}$ edge was fitted with a superposition of $3d^{3+x} L^x$ configurations. (B) $3d^6$ configuration of Co, fitted with CTM4XAS multiplet calculations.

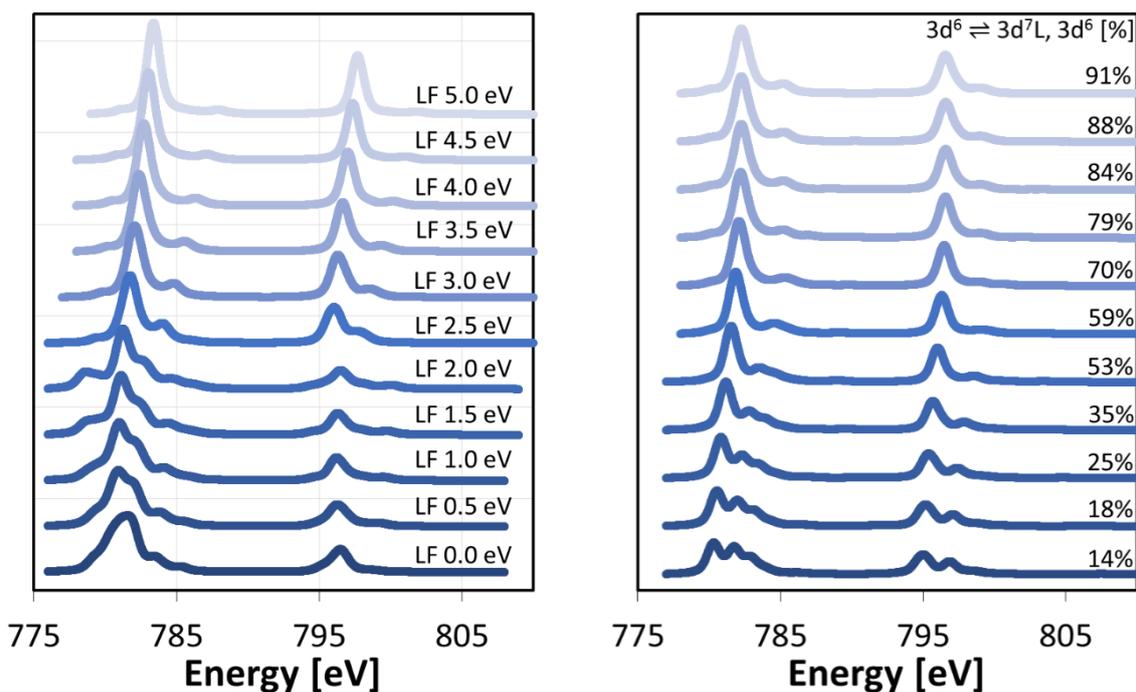

**Figure S7:** CTM4XAS calculations of $Co^{3+}$ with a variation in the Ligand-field (A) (LF energy was varied from 0.0 eV to 5.0 eV in 0.5 steps) and a variation in the charge-transfer-energy (B) (LF = 3 eV, LMCT was varied from -10 eV to -10 eV in 2 eV steps).

## S5. Fits of the O K edges

The O K edges of NCM 111, NCM 622 and NCM 811 are fitted with three Gaussian functions in order to depict main changes between the charged and the discharged samples. Regardless of the many *Me*-O states leading to the peaks in the O K NCM spectra (see **Figure 5**, main text), the O K edge pre-peaks can be sufficiently fitted with three Gaussian functions (**Figure S8**). Gauss1 and Gauss2 increase with an increasing state of charge as the $Ni^{3+}$ content does (**Figure S8**, comparison of first and second column). The $Ni^{3+}$ content is also increasing with an increasing Ni content (**Figure S5**, discharged samples) but only Gauss1 is following this trend (**Figure S8**, discharged NCM 111, NCM 622 and NCM 811). In discharged NCM 111, NCM 622 and NCM 811 the area of Gauss2 is similar. Since the $Ni^{3+}$ content is increasing significantly with a higher Ni content in the samples, other transition metal hybrid orbitals such as Co-O and Mn-O, which decrease with an increasing Ni-content, also contribute to this peak as already discussed in the main text. The $Ni^{2+}$ content does only slightly increase with an increasing Ni content and vanishes upon charge. Gauss 1 and Gauss 2 are still present in the charged spectra and therefore $Ni^{2+}$ cannot explain the presence of the peaks. Moreover, the O K NEXAFS spectra of $Ni^{2+}$ (e.g. in NiO [9]) do not show peaks between 528 eV and 530 eV, the energy region where Gauss1 and Gauss2 are present. Gauss3 does not change upon charge/discharge but decreases from NCM 111 to NCM 811 (with an increasing Ni content). This can be explained by the decreasing Mn and Co content in the NCMs, which are electrochemically inactive transition metals and show *Me*-O peaks between 529 eV and 535 eV (Peak D and E, **Figure 5**, main text). However, it should be noted that possible peaks due to excitations of electrons into Hubbard bands are also lying in this energy range.[10,11]

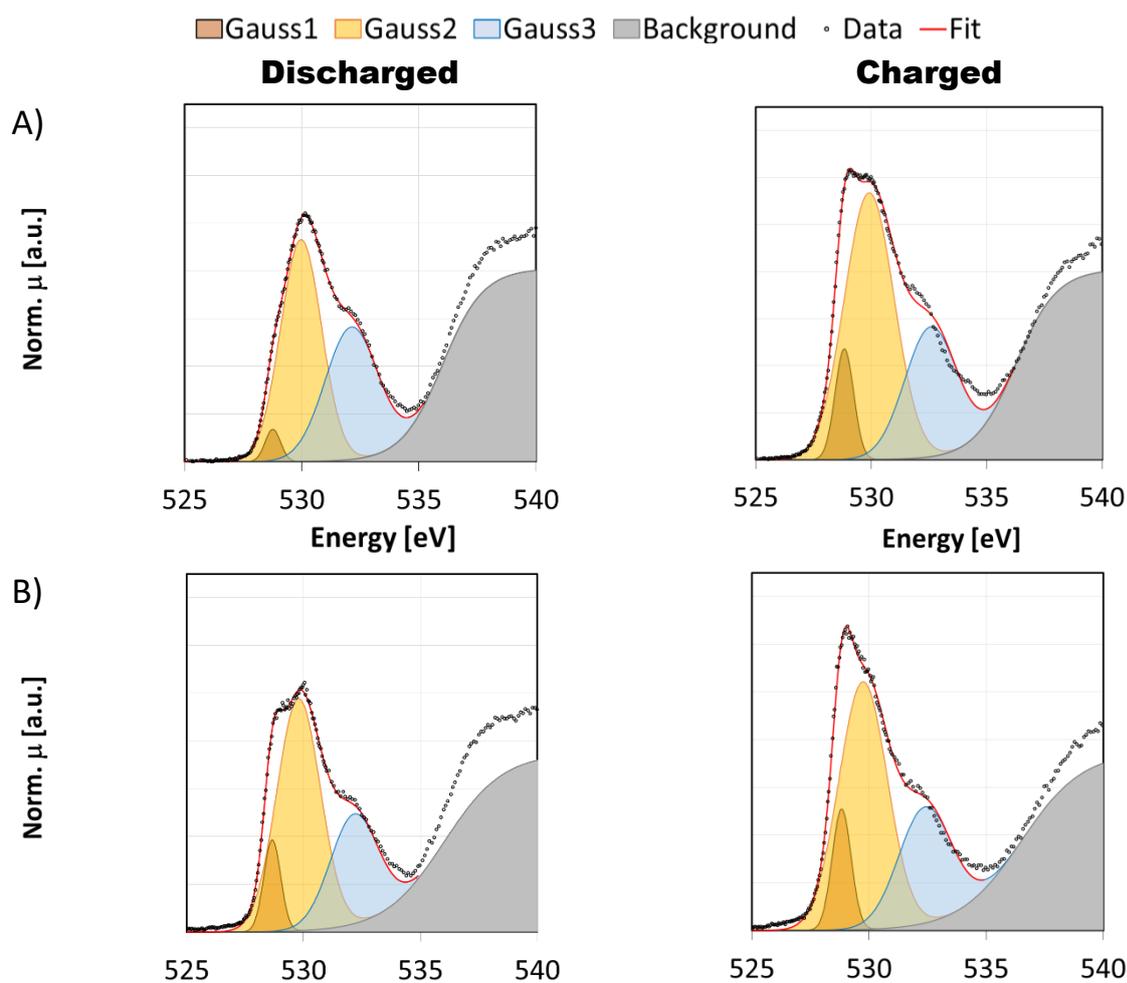

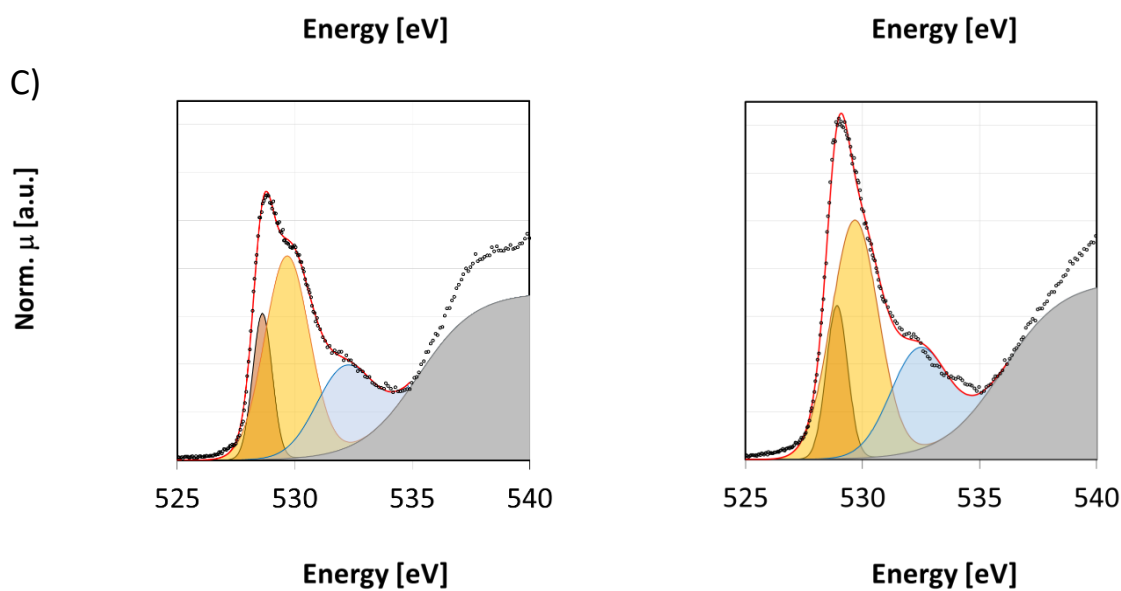

**Figure S8**: Fit of the O K XAS spectra of NCM 111 (A), NCM 622 (B) and NCM 811 (C), measured in the discharged (1$^{st}$ column) and charged (2$^{nd}$ column) state of the materials. The data is fitted with three gaussian functions (gauss1: orange, gauss2: yellow, gauss3: blue). The grey area is the background of the spectra normalized.

## S6. Synthesis and characterization of the NCMs, LNO, LCO and LLO

- NCMs

The NCMs were synthesized using a precipitation method. Therefore, a stoichiometric amount of NiSO$_4$· 6 H$_2$O (Sigma Aldrich, > 98%), CoSO$_4$· 7 H$_2$O (Sigma Aldrich, > 99%) and MnSO$_4$· H$_2$O (Sigma Aldrich, > 99%) were dissolved in distilled water (2 mol/L).[12] N$_2$ gas was bubbled through while stirring the solution in a tank reactor and the temperature was adjusted to 60°C. NH$_4$OH (28% NH$_3$ in water, Sigma Aldrich, > 99%, NH$_4$OH:Me = 1.2:1 with Me = Ni, Co, Mn) was added at a rate of 0.2 ml min$^{-1}$ to the solution (atomic ratio of 1.2 to the Me-salts). The pH was continuously adjusted to 11 by further adding NaOH (Sigma Aldrich, > 99%). The solution was stirred for 12 h at 50°C while the pH was kept at 11 by further adding NaOH. The

---

precipitation was filtered and dried at 120°C. Together with LiOH·H$_2$O (Sigma Aldrich, > 98%, Li:Me = 1.1:1) the precipitation was first heated to 400°C, hold for 5 h (pre-calcination), than heated to 800°C and hold under oxygen for 10 h to obtain the layered oxides.

- Reference materials

The layered oxides LiCoO$_2$ and LiNiO$_2$ were synthesized with a water-based sol gel method. Co(CH$_3$COO)$_2$·4H$_2$O (Merckmillipore, > 99.0 %) or Ni(CH$_3$COO)$_2$·4 H$_2$O (Sigma Aldrich, > 99.0 %) and Li(CH$_3$COO) (Sigma Aldrich, 99,99 %) in an atom ratio of 1:1.1-1.2 were dissolved in water (0.2 molar Li/Me solution) while citric acid (CA, Sigma Aldrich, >99.5 %) was added as a chelating agent in an atom ratio of 1 to the transition metals (Me). The pH was corrected with ammonium hydroxide as given in references[13,14]. Afterwards the solution was heated up to 80°C for gelation and the temperature was held until the obtained gel was completely dry. To ensure the homogeneity of the powders during thermal treatment, the dry gel was ground in an agate mortar before annealing. In the case of Co rich layered oxides, the annealing was performed with a heating rate of 1 K/min up to 400 °C followed by a 1 h holding step to remove all carbon residuals.[15,16] Then, the samples were heated to 700°C with 5 K/min and were hold at this temperature for 6 h. The heat treatment was performed in a static synthetic air atmosphere. Based on TGA/MS measurements the heat treatment for LNO was performed in 2 steps. The pre-calcination step of the precursors was done in air at 600°C for 6 h. Subsequently, the particles were ground and calcined at 800°C for 13h in an oxygen atmosphere. [17,18] LLO was synthesized using a solid-state method.[19] Stoichiometric amounts of Li$_2$CO$_3$ (Sigma-Adrich®, ≥ 99%) and MnO$_2$ (Alfa-Aesar®, 99.9%) with a molar ratio of 2:1 (Li:Mn) were mixed in a planetary ball mill using YSZ-balls and isopropanol as dispersion agent. The solvent was removed and the powder mixture was calcined in static air for 48 h at 800 °C in Al$_2$O$_3$ crucibles with a heating rate of 5 K/min.

- Powder diffraction

Layered oxides (NCMs, LNO and LCO) crystallize in the space group $R\bar{3}m$ in which Li ions and transition metals occupy edge-sharing oxygen octahedra in alternating layers, **Figure S9A**. In LLO, excess Li ions replace some of the transition metal ions forming a honey comb ordering in the transition metal planes, breaking the rhombohedral symmetry down to a monoclinic $C\,2/m$ structure as shown in **Figure S9B** .The starting parameters for the refinements are given in **Table S2**.

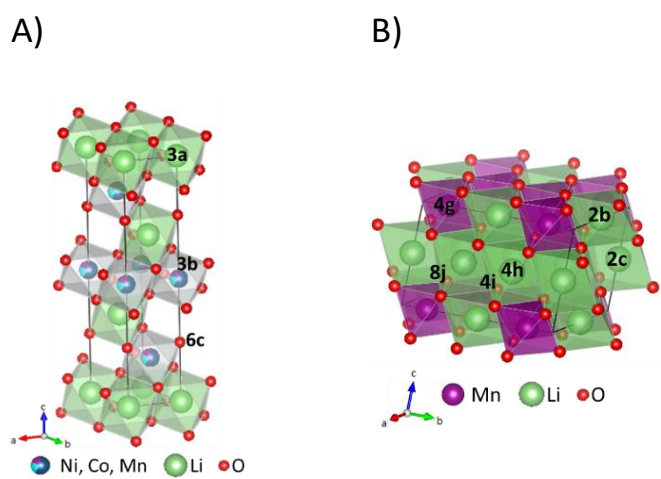

**Figure S9:** 3D illustration of the rhombohedral $\boldsymbol{R\bar{3}m}$ of layered oxides (A) and the monoclinic $\boldsymbol{C\,2/m}$ unit cell of Li$_2$MnO$_3$ (B).

**Table S2:** Structural starting parameters used for the refinement of the rhombohedral A) and monoclinic phase B).

A)

| sites | element | x | y | z | Occupancy |
|---|---|---|---|---|---|
| **3a** | Li | 0.00000 | 0.00000 | 0.50000 | 0.0833(4) |
| **3b** | Ni, Co, Mn | 0.00000 | 0.00000 | 0.00000 | 0.0833(2) |
| **6c** | O | 0.00000 | 0.00000 | 0.2564(3) | 0.16667 |

B)

| sites | element | x | y | z | Occupancy |
|-------|---------|---------|---------|---------|-----------|
| 4h | Li | 0.00000 | 0.33940 | 0.50000 | 0.1111(4) |
| 2c | Li | 0.00000 | 0.00000 | 0.50000 | 0.0557(2) |
| 2b | Li | 0.00000 | 0.00000 | 0.00000 | 0.0557(2) |
| 4g | Mn | 0.00000 | 0.00000 | 0.16708 | 0.1111(2) |
| 8j | O | 0.25400 | 0.32119 | 0.26004 | 0.22222 |
| 4i | O | 0.21890 | 0.00000 | 0.26004 | 0.11111 |

**Figure S10** shows the powder diffraction patterns of NCM 111, NCM 622, NCM 811, LNO, LCO and LLO collected at beamline I11 with the high-resolution MAC detector [20] (can probe impurities with a phase content <= 2%) and the corresponding refinements to confirm that the obtained powders are phase pure. The energy of the X-Ray beam was tuned to 15 keV and the calibrated wavelength was 0.825270(10) Å. The pristine materials were measured in a borosilicate capillary (0.5 mm outer diameter) and the 2D data was rebind to a step size of 0.005° 2θ using a Diamond software. The refinement was performed as described in the experimental section of the main text. The input parameters for the reference materials were obtained from crystallographic data files (NCM 111[21], NCM 622 [22], NCM 811 [23], LNO [24], LCO [25], LLO [26]) and the results of the refinements of the pristine references are given in **Table S3**.

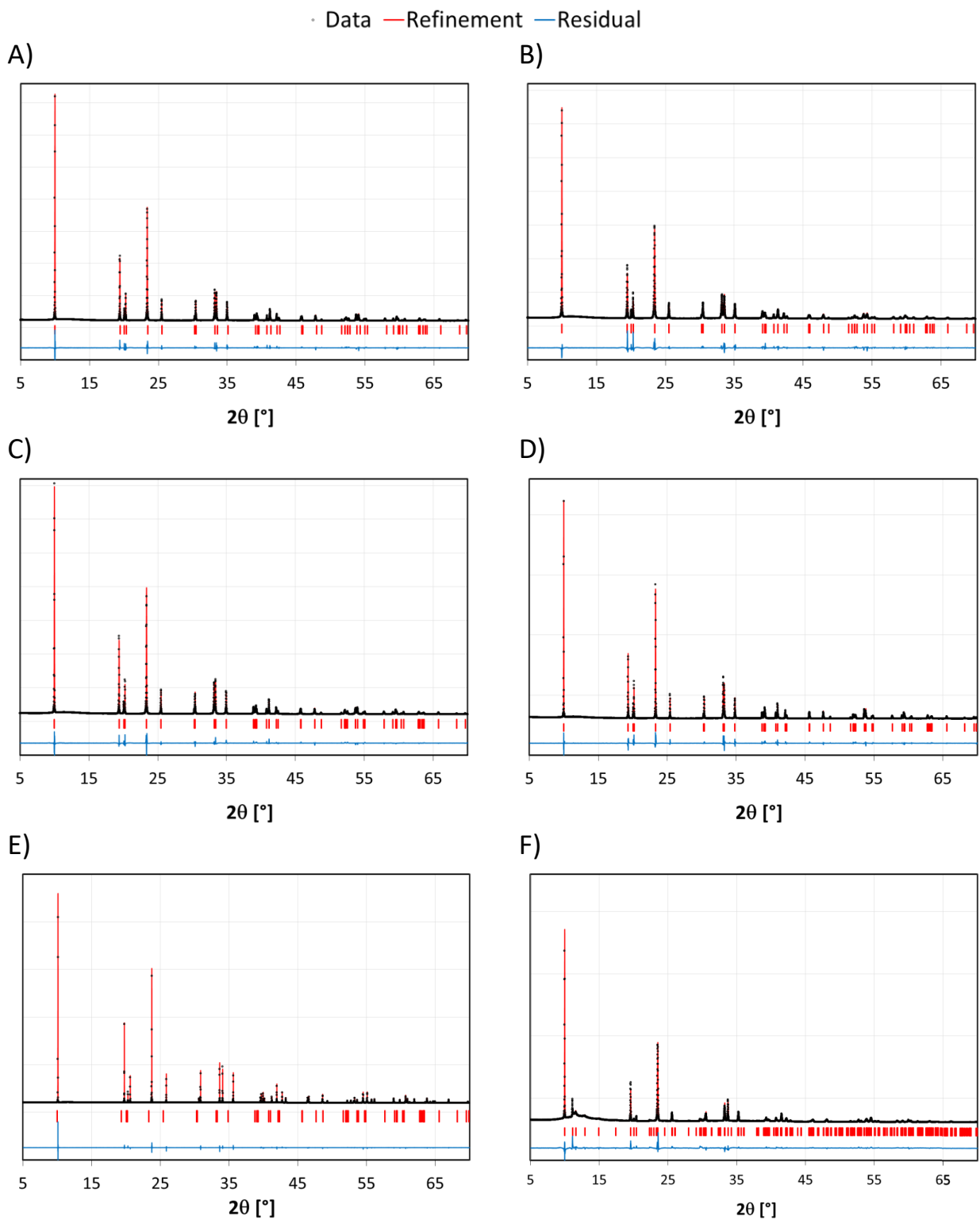

**Figure S10:** Powder diffraction pattern and refinement of NCM 111 (A), NCM 622 (B), NCM 811 (C), LNO (D), LCO (E) and LLO (F), collected at beamline I11 with the MAC detector.

**Table S3:** Results of the refinement of LNO (A), LCO (B) and LLO (C). In the case of LNO Li/Ni disorder was refined in addition.

A)

| $R\bar{3}m$ | Refinement |
|---|---|
| Lattice param. a, b | 2.85770(1) Å |
| Lattice param. c | 14.2215(1) Å |
| Density | 4.950(1) g cm-3 |
| Volume | 100.579(1) Å3 |
| Li/Ni disorder | 3.2% |
| $R_{Bragg}$ | 2.98 |

B)

| $R\bar{3}m$ | Refinement |
|---|---|
| Lattice param. a, b | 2.868586(5) Å |
| Lattice param. c | 14.21543(4) Å |
| Density | 4.751(1) g cm$^{-3}$ |
| Volume | 101.304(1) Å$^3$ |
| Li/Ni disorder | 2.2% |
| $R_{Bragg}$ | 2.60 |

C)

| $R\bar{3}m$ | Refinement |
|---|---|
| Lattice param. a, b | 2.871949(5) Å |
| Lattice param. c | 14.20085(4) Å |
| Density | 4.711(1) g cm$^{-3}$ |
| Volume | 101.437(1) Å$^3$ |
| Li/Ni disorder | 1.3% |
| $R_{Bragg}$ | 2.55 |

D)

| $R\bar{3}m$ | Refinement |
|---|---|
| Lattice param. a, b | 2.881718(8) Å |
| Lattice param. c | 14.21208(7) Å |
| Density | 4.349(1) g cm$^{-3}$ |
| Volume | 102.210(1) Å$^3$ |
| Li/Ni disorder | 5.4% |
| $R_{Bragg}$ | 5.90 |

E)

| $R\bar{3}m$ | Refinement |
|---|---|
| Lattice param. *a, b* | 2.81747(4) Å |
| Lattice param. *c* | 14.06200(3) Å |
| Density | 4.749(1) g cm$^{-3}$ |
| Volume | 96.671(1) Å$^3$ |
| Li/Ni disorder | N/A |
| $R_{Bragg}$ | 4.68 |
| | |

F)

| $R\bar{3}m$ | Refinement |
|---|---|
| Lattice parameter *a* | 4.92759(6) Å |
| Lattice parameter *b* | 8.5273(1) Å |
| Lattice parameter *c* | 5.02241(7) Å |
| β angle | 109.301(2) |
| Volume | 199.289(5) Å$^3$ |
| Density | 6.221(1) g cm$^{-3}$ |
| $R_{Bragg}$ | 6.265 |

- SEM images of the NCMs

Scanning electron microscopy (SEM) pictures of NCM 111 (**Figure S11 A, B**), NCM 622 (**Figure S11 C, D**) and NCM 811 (**Figure S11 E, F**) were obtained using a Hitachi TM-1000 table top instrument.

A)

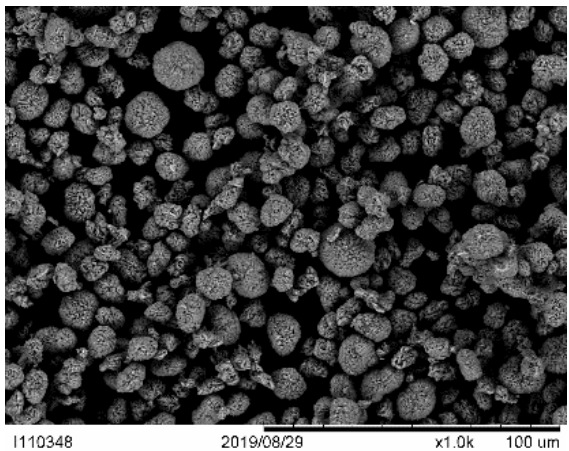

B)

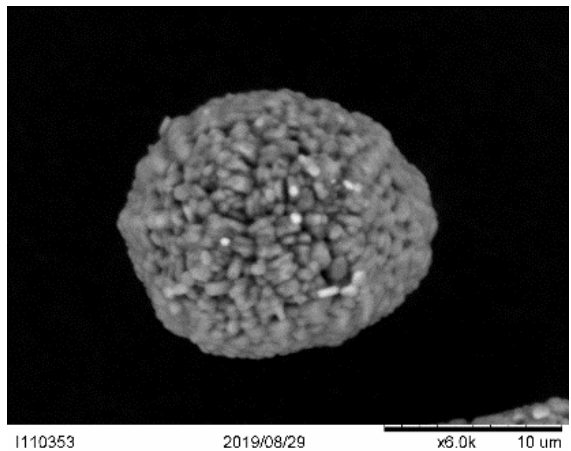

C)

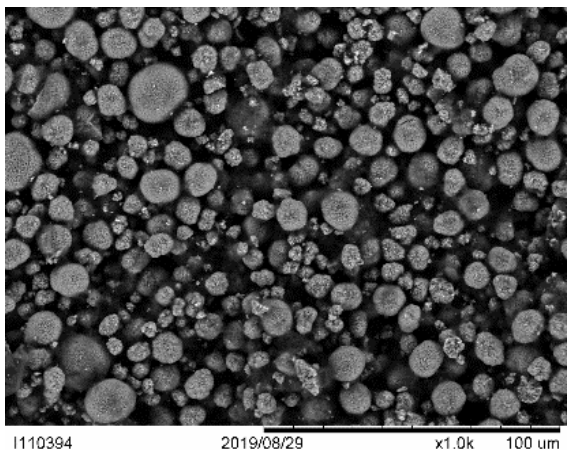

D)

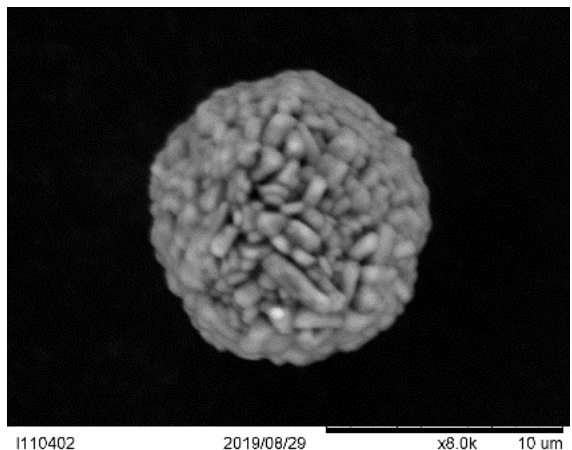

E)

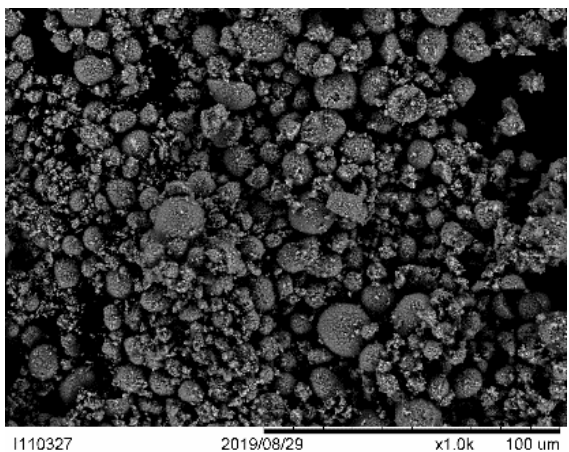

F)

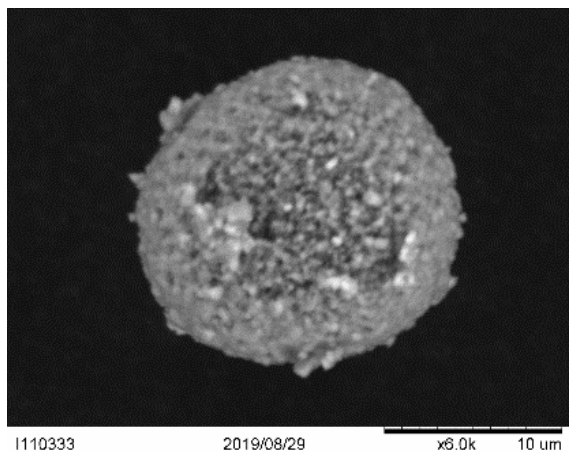

**Figure S11:** SEM images of NCM 111 (A) & (B), NCM 622 (B) & (C) and NCM 811 (D) &

(E).

## S7. Impact of electrolyte decomposition products and binder on XAS spectra

Decomposition products deposited at the cathode/electrolyte interface are mainly Li salts like carbonates, fluorides or phosphates.[27] Carbonates and phosphates lead to additional peaks in the O K spectra while the F K edge appears at ca. 698 eV with EXAFS features at higher energies interfering with the Co $L$ and Ni $L$ edge. Fluorine is also present in the materials due to PvdF (binder, see experimental section) used to coat the active materials. An overview spectrum of a NCM 111 (100 cycles, cycled with C/2) before and after rinsing with dimethyl carbonate (DMC) for 5 min is provided for three reasons: (i) To ensure that the O K edge peaks in the FY spectra are correctly attributed to either the bulk material or to electrolyte decomposition products, (ii) to prove to which extent the Co $L$ and Ni $L$ edge are affected by the F K edge EXAFS features and (iii) to figure out, whether rinsing can help to minimize these effects (**Figure S12**). NCM 111 was cycled between 2.0 V and 4.6 V to ensure that electrolyte decomposition reactions take place and cathode surface reactions and transition metal dissolution become more pronounced. The F K edge dominates the overview spectra showing that fluorine (stemming from electrolyte decomposition products or binder) is the main component present on the cathode/electrolyte interface. It also becomes obvious that rinsing the sample with DMC does not reduce the amount of fluorine since the intensity of the F K edge does not decrease.

---

27      Edström et al., **2004**, *50* (2-3), 397.

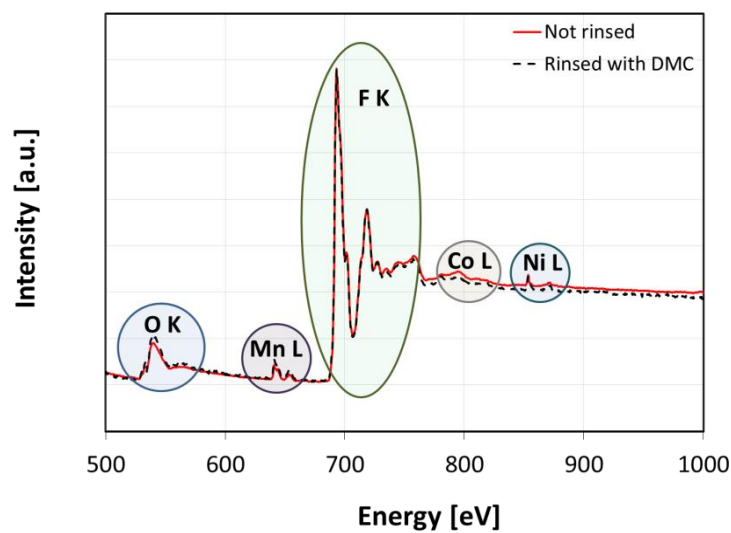

**Figure S12:** Overview spectra of NCM 111 (TEY signal/$I_0$), cycled for 100 samples (black dashed line) and rinsed with DMC for 5 min (red dotted line).

In order to reveal the effect of decomposition products on the O K spectra, **Figure S12** shows the FY O K spectra of pristine NCM 111, NCM 111 after 100 cycles as well as NCM 111 after 100 cycles and rinsed for 5 min with DMC. No additional peaks are found comparing pristine NCM 111 (which was not exposed to electrolyte) with NCM 111 after 100 cycles, no matter if the sample was rinsed or not. The absence of peaks at 533 eV and 535 eV (O K peak of LiNiPO$_4$ and Li$_2$CO$_3$ [28,29], the salts are expected to be the main electrolyte decomposition products on the cathode surface) further confirms that the effect of surface impurities can be neglected considering FY spectra.

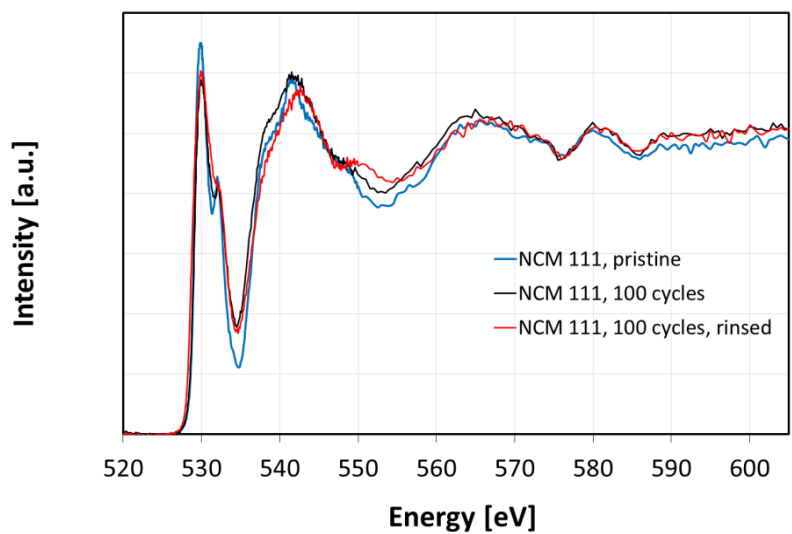

**Figure S13:** FY O K XAS of pristine NCM 111 (blue line), NCM 111 after 100 cycles (black line) and NCM 111 after 100 cycles and rinsed with DMC for 5 min in an argon filled glovebox.

**Figure S14** shows FY spectra of the Ni $L_{2,3}$ (A) and Co $L_{2,3}$ (B) edge of NCM 111 after 100 cycles (black: without rinsing, red: with rinsing). As discussed, rinsing does not lead to significant changes. Nevertheless, it has to be mentioned that the EXAFS features of the F K edge overlap significantly with the Co $L_{2,3}$ and to a minor extent with the Ni $L_{2,3}$ edge leading to difficulties in the background correction of the spectra. Therefore, the slope of the background was determined using an energy range after the $L_{2,3}$ edges. The line was then positioned into the minimum before the edge jump (average of 5 points around the minimum in intensity before the onset of the $L_3$ edge).

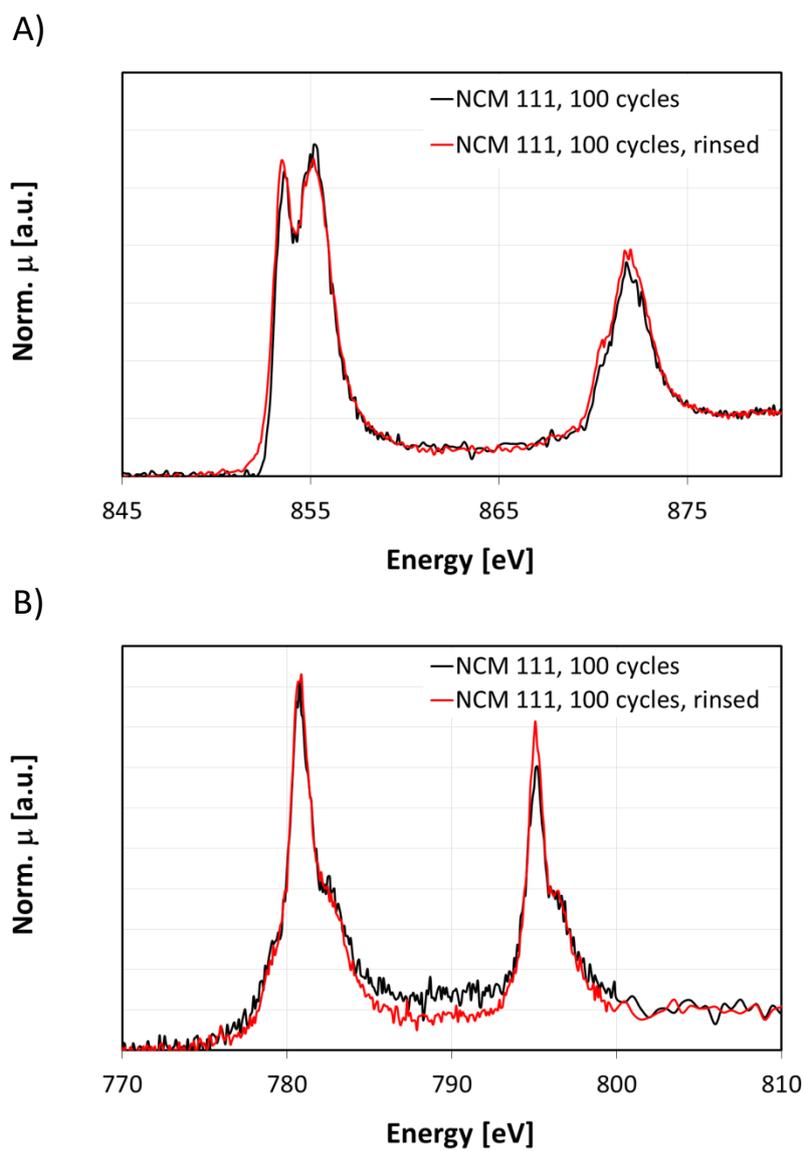

**Figure S14:** (A) Ni $L_{2,3}$ and (B) Co $L_{2,3}$ edge of NCM 111 after 100 cycles (black lines) as well as after rinsing.